\documentstyle[prd,aps,preprint,epsf,axodraw]{revtex}
\newcommand{\vev}[1]{ \left\langle {#1} \right\rangle }
\newcommand{\barr}[1]{ \overline{{#1}} }

\newcommand{\GeV}{~\mbox{GeV}}

\newcommand{\kahler}{K\"ahler }

\def\SO{\mathop{\rm SO}}
\def\O{\mathop{\rm O}}
\def\SU{\mathop{\rm SU}}
\def\U{\mathop{\rm U}}
\def\Sp{\mathop{\rm Sp}}
\def\SL{\mathop{\rm SL}}

\def\diag{\mathop{\rm diag}\nolimits}

\def\Re{\mathop{\rm Re}\nolimits}
\def\Im{\mathop{\rm Im}\nolimits}
\begin{document}
\tighten
\draft

 \baselineskip 0.7cm
\renewcommand{\thefootnote}{\fnsymbol{footnote}}
\setcounter{footnote}{1}
 
\title{\hfill{\normalsize\vbox{\hbox{UT-928 }
}}\\
Semi-simple group unification \\ in the 
          supersymmetric brane world}
 \vskip 1.2cm
\author{Y. Imamura and  T. Watari}
\address{Department of Physics, University of Tokyo, Tokyo 113-0033,
Japan}
\author{T. Yanagida}
\address{Department of Physics, University of Tokyo, Tokyo 113-0033,
Japan \\ and \\
Research Center for the Early Universe, University of
Tokyo, Tokyo, 113-0033, Japan}

\date{\today}
\maketitle

\vskip 1cm
\begin{abstract}
The conventional supersymmetric grand unified theories suffer from two
serious problems, the large mass splitting between doublet and triplet 
Higgs multiplets, and the too long lifetime of the proton. A unification 
model based on a semi-simple group SU(5)$_{\rm GUT}\times$U(3)$_{\rm H}$ 
has been proposed to solve both of the problems simultaneously. 
Although the proposed model is perfectly consistent with observations, 
there are various mysteries. In this paper, we show that such mysterious 
features in the original model are naturally explained by embedding 
the model into the brane world in a higher dimensional space-time. 
In particular, the relatively small gauge coupling constant of the 
SU(5)$_{\rm GUT}$ at the unification energy scale is a consequence of 
relatively large volume of extra dimensions. Here, we put the 
$\SU(5)_{\rm GUT}$ gauge multiplet in a 6-dimensional bulk and assume 
all fields in the $\U(3)_{\rm H}$ sector to reside on a 3-dimensional 
brane located in the bulk.  On the other hand, all chiral multiplets 
of quarks, leptons and Higgs are assumed to reside on a 3-brane at a 
${\bf T}^2/{\bf Z}_4$ orbifold fixed point. The quasi-${\cal N}=2$ 
supersymmetry 
in the hypercolor U(3)$_{\rm H}$ sector is understood as a low-energy 
remnant of the ${\cal N}=4$ supersymmetry in a 6-dimensional
space-time. 
We further extend the 6-dimensional model to a 10-dimensional theory. 
Possible frameworks of string theories are also investigated 
to accommodate the present brane-world model. We find that the type IIB 
string theory with D3-D7 brane structure is an interesting candidate. 
\end{abstract}


\renewcommand{\thefootnote}{\arabic{footnote}}
\setcounter{footnote}{0}

\section{Introduction}
\label{sec:introduction}
%
%
The supersymmetry (SUSY) is a very interesting symmetry which provides a
natural explanation of the light Higgs scalar doublet in the standard
model. That is, the SUSY-invariant mass term for the Higgs chiral
multiplets can be suppressed by an appropriate chiral symmetry in the SUSY 
standard model, and if the breaking of the chiral symmetry is linked to
the SUSY breaking, the masses of Higgs scalar doublets are naturally
predicted at the SUSY-breaking scale ($\sim 1$ TeV) \cite{masiero}. 
However, we have to abandon this beautiful Giudice-Masiero mechanism
\cite{masiero} in SUSY grand unified theories (GUT's), since the Higgs 
doublets are necessarily accompanied by color-triplet Higgs multiplets 
whose masses should be at least of the order of unification scale 
$\sim 10^{16}$ GeV to account for the observed stability of proton. 
The chiral symmetry that forbids the SUSY-invariant mass term for the 
Higgs doublets also forbids the mass term for the Higgs color-triplets 
and their masses are also predicted at the SUSY-breaking scale inducing
too rapid proton decay.

A SUSY unification model based on a semi-simple gauge group 
SU(5)$_{\rm GUT}\times$U(3)$_{\rm H}$ has been proposed to solve the 
above problem \cite{yanagida,hy,iy}, in which the color-triplet Higgs 
multiplets acquire SUSY-invariant masses together with newly introduced 
colored chiral multiplets while the Higgs doublets remain massless. 
As a direct consequence of the Higgs structure this model solves also 
another problem in the SUSY standard GUT's; that is, the dangerous
dimension-five operators \cite{syw} for the proton decay are suppressed
by the symmetry that forbids the mass term for Higgs doublets. In this 
model the low-energy color 
group SU(3)$_{\rm C}$ is a diagonal subgroup of SU(3)$\times$SU(3)$_{\rm 
H}$ where the first SU(3) is a subgroup of the SU(5)$_{\rm GUT}$. The 
remarkable feature in this model is that the gauge coupling constants of 
the hypercolor group U(3)$_{\rm H}$ should be very large, while the gauge
coupling of the SU(5)$_{\rm GUT}$ is relatively small at the unification 
scale $M_{\rm GUT}$, so that we realize the approximate unification 
of three gauge coupling constants of the low-energy standard-model gauge 
group SU(3)$_{\rm C}\times$SU(2)$_{\rm L}\times$U(1)$_{\rm Y}$. 

In this paper 
we embed this semi-simple unification model into the brane world in a
higher dimensional space-time and show that the disparity of gauge 
coupling strengths is naturally understood in terms of relatively 
large volume 
of the extra dimensions. We put the SU(5)$_{\rm GUT}$ gauge vector 
multiplet in a 6-dimensional bulk and assume that the extra 
2-dimensional space is compactified on an orbifold ${\bf T}^2/{\bf Z}_4$. 
The standard 
quark, lepton and Higgs chiral multiplets are assumed to reside on a 
3-dimensional brane (3-brane) at one of the orbifold fixed points. 
We see that the SU(5)$_{\rm GUT}$ gauge coupling receives a volume 
suppression due to the extra dimensions. On the other hand, we assume
that the hypercolor U(3)$_{\rm H}$ sector resides on a 3-brane located 
in the bulk (not at the fixed points). Thus, the gauge interactions of
U(3)$_{\rm H}$ are no longer suppressed. This configuration may also account 
for another mystery in the original  SU(5)$_{\rm GUT}\times$U(3)$_{\rm
H}$ model, namely the ${\cal N}=2$ SUSY structure of the hypercolor 
U(3)$_{\rm H}$ sector.
 
In our brane world the fundamental scale is determined as $M_* \simeq
10^{17}$ GeV for a successful phenomenology and the size of the 
compactified space is fixed as $\sim 10^{16}$ GeV so that 
the 4-dimensional Planck scale $M_{\rm Pl}\simeq 2\times 10^{18}$ GeV
is obtained \cite{witten,nima}. This lower-energy fundamental scale 
provides a solution to a potential problem in the original model that 
the gauge coupling constant of U(1)$_{\rm H}$ is asymptotic non-free and 
it blows up  around $6\times 10^{17}$ GeV below the Planck scale. 
However, this is not a problem in our brane world scenario, since the 
4-dimensional Planck scale is merely an effective one and the present
model is considered as only a low-energy description below the fundamental 
scale $M_*\simeq 10^{17}$ GeV of a more fundamental theory. 

It is very attractive to consider that the SUSY is broken on a hidden   
3-brane at another orbifold fixed point \cite{randall}. If it is the case, 
the present model is an extension of the gaugino-mediation model 
\cite{kawasaki,kaplan} of the SUSY breaking and provides a natural 
solution to the SUSY flavor problem. Here, the SU(5)$_{\rm GUT}$ gaugino
acquires a SUSY-breaking mass since they live in the 6-dimensional bulk 
and couples directly to the hidden-sector field. On the contrary, the
U(3)$_{\rm H}$ gauginos remain massless at the tree level, since they
are localized on the 3-brane separated from the SUSY-breaking hidden
brane.  In this case we find an approximate GUT-unification of the
masses of the  SUSY standard-model gauginos, $m_{\widetilde{G}_3}\simeq
m_{\widetilde{G}_2}\simeq m_{\widetilde{G}_1}$, at the unification scale 
$M_{\rm GUT}$. This result is
very interesting, because the gaugino masses can be different from each
others in the original semi-simple unification model \cite{moroi}.

In section \ref{sec:5-3}, we review briefly the SUSY $\SU(5)_{\rm
GUT} \times \U(3)_{\rm H}$ unification model, and explain why the
doublet-triplet mass splitting problem for Higgs multiplets is naturally 
solved. Here, we point out that
there are various mysterious, but interesting features in the original
model that are required for successful phenomenologies. 
In section \ref{sec:embedding}, we embed the  
$\SU(5)_{\rm GUT} \times \U(3)_{\rm H}$ in the brane world in a higher
dimensional space-time. We explain,
here, why we choose the dimension of extra space to be 2 and why the orbifold 
 compactification is necessary. We consider the ${\bf T}^2/{\bf Z}_4$
orbifold as an example. We find that ${\cal N}=4$
SUSY in the 6-dimensional bulk is crucial to have ${\cal N}=2$ SUSY on
the U(3)$_{\rm H}$ 3-brane. We show here that mysterious features in the 
original model are indeed naturally explained by the present
embedding of the original model into the 6-dimensional space-time.
In section \ref{sec:gaugino-med}, we discuss 
SUSY-breaking effects provided that the SUSY is broken on a hidden 3-brane
at an orbifold fixed point. We discuss, in section
\ref{sec:string},
 a possible 
connection to string theories, since we consider that the string
theories provide a natural framework of the brane world in a higher
dimensional space-time. We find a preferable scheme may be
provided by the type IIB string theory with D3-D7 branes. However, we also
note that there are various unsolved problems in this string framework
which may deserve further investigations. 
The last section is devoted to discussion and conclusions.


\section{SUSY SU(5)$_{\rm GUT}\times$U(3)$_{\rm H}$ unification model}
\label{sec:5-3}
%
%
In this section, we discuss briefly a semi-simple unification model 
based on an ${\cal N}=1$ SUSY SU(5)$_{\rm GUT}\times $U(3)$_{\rm H}$ 
gauge theory \cite{yanagida,hy,iy}. The 
SU(5)$_{\rm GUT}$ is the usual GUT gauge group and its gauge coupling 
constant is in a perturbative regime, $\alpha _{\rm GUT}\simeq 1/24$. 
The U(3)$_{\rm H}$ is a hypercolor gauge group whose gauge interactions 
are sufficiently strong at the unification scale $M_{\rm GUT}$ as shown
below. The usual quark and lepton chiral multiplets transform as ${\bf
5^*}$ and ${\bf 10}$ under the SU(5)$_{\rm GUT}$ and they are all
singlets of the hypercolor U(3)$_{\rm H}$. A pair of Higgs multiplets
$H_k$ and ${\bar H}^k (k=1,...,5)$ transforms as ${\bf 5+5^*}$ under the
SU(5)$_{\rm GUT}$ and as singlets under the U(3)$_{\rm H}$.

In addition to the usual matter chiral multiplets we introduce six pairs 
of hyperquarks $Q^\rho_\alpha$ and ${\bar Q}^\alpha _\rho$ $~(\alpha =1,2,3;
\rho =1,...,6)$ which belong to ${\bf 3 + 3^*}$ of the hypercolor
SU(3)$_{\rm H}$
and have U(1)$_{\rm H}$ charges $1$ and $-1$, respectively. Here,
$\U(3)_{\rm H}\equiv \SU(3)_{\rm H}\times \U(1)_{\rm H}$. The first 
five pairs of $Q^k_\alpha$ 
and ${\bar Q}^\alpha _k (\alpha =1,2,3; k=1,...,5)$ transform as ${\bf
5^*+5}$ under the  SU(5)$_{\rm GUT}$ and the last pair of $Q^6_\alpha$ 
and ${\bar Q}^\alpha _6$ is a singlet of the  SU(5)$_{\rm GUT}$. To cause 
the desired breaking of the total gauge group SU(5)$_{\rm
GUT}\times$U(3)$_{\rm H}$ down to the standard-model gauge group
SU(3)$_{\rm C}\times$SU(2)$_{\rm L}\times$U(1)$_{\rm Y}$, we furthermore 
introduce chiral multiplets $X^\alpha _\beta$ and $X_0$ which are an 
adjoint and a singlet representation of the $\SU(3)_{\rm H}$, respectively 
\cite{iy}. They would
be regarded as ${\cal N}=2$ SUSY partners of the vector multiplets of 
the hypercolor U(3)$_{\rm H}$ as seen in the next section.

We now introduce a superpotential,
\begin{equation}
W = \lambda {\bar Q}^\beta _kX^\alpha _\beta Q^k_\alpha   
  +  \lambda '{\bar Q}^\beta _6X^\alpha _\beta Q^6_\alpha 
  + \kappa  {\bar Q}^\alpha _kX_0 Q^k_\alpha 
  + \kappa '{\bar Q}^\alpha _6X_0 Q^6_\alpha 
  - 3V^2X_0.
\label{superpotential}
\end{equation}
We have the ${\cal N}=2$ SUSY \cite{hy} in the hypercolor sector in 
the limit of $\lambda =\lambda '=\sqrt{2}g_{3{\rm H}}$ and 
$\kappa =\kappa '=\sqrt{2}g_{1{\rm H}}$, where $g_{3{\rm H}}$ and
$g_{1{\rm H}}$ are gauge 
coupling constants of the SU(3)$_{\rm H}$ and the U(1)$_{\rm H}$, 
respectively. We note here that the last term in eq.(\ref
{superpotential}) corresponds to the Feyet-Iliopoulos (FI) F-term 
\cite{fi-term} and it is perfectly allowed by the ${\cal N}=2$ SUSY. 
However, even if such an ${\cal N}=2$ SUSY relation holds at the classical
level, the  ${\cal N}=2$ SUSY in the hypercolor sector is explicitly
broken by interactions with other sectors, since the SUSY of the total
system is only ${\cal N}=1$. For instance, quantum corrections from the 
SU(5)$_{\rm GUT}$ gauge interactions change various coupling constants 
in the hypercolor U(3)$_{\rm H}$ sector differently, 
loosing the ${\cal N}=2$ relation 
among the Yukawa and gauge coupling constants. Thus, we do not impose 
such a restrictive ${\cal N}=2$ SUSY condition in this paper. We see that 
the ${\cal N}=2$ SUSY relation of coupling constants is not necessarily
crucial for the following hypercolor dynamics. 

As shown in ref.~\cite{iy} we have a desired vacuum,
\begin{equation}
\langle Q^r_\alpha \rangle = \frac{V}{\sqrt{\kappa}}\delta ^r_\alpha ,~~
\langle {\bar Q}^\alpha_r \rangle = \frac{V}{\sqrt{\kappa}}\delta _r^\alpha ,~~
\langle X^\alpha _\beta \rangle = \langle X_0 \rangle =
\langle  Q^6_\alpha \rangle = \langle {\bar Q}^\alpha_6 \rangle = 0.
\label{vacuum}
\end{equation}
Notice that this classical vacuum exists even at the quantum level
\cite{seiberg,iy}. It may be instructive to know that this vacuum is in a
Higgs branch in an ${\cal N}=2$ SUSY QCD \cite{seiberg2} that may be
stable against the present deviation from the ${\cal N}=2$ to the 
${\cal N}=1$ SUSY. 
In this vacuum with $V\ne 0$ the total gauge group SU(5)$_{\rm
GUT}\times$U(3)$_{\rm H}$ is broken down to the SU(3)$_{\rm
C}\times$SU(2)$_{\rm L}\times$U(1)$_{\rm Y}$ and thus we take the 
vacuum-expectation value (vev) to be the unification scale $V \simeq M_{\rm
GUT} \simeq 10^{16}$ GeV. Here, the color SU(3)$_{\rm C}$ is 
an unbroken diagonal subgroup of 
the SU(3)$\times$SU(3)$_{\rm H}$ where the first SU(3) is a subgroup of
SU(5)$_{\rm GUT}$, and the U(1)$_{\rm Y}$ is a linear combination of a
U(1) subgroup of SU(5)$_{\rm GUT}$ and the hypercolor U(1)$_{\rm
H}$. Thus, the gauge coupling constants $\alpha _3, \alpha _2$ and
$\alpha _1$ of the low-energy 
SU(3)$_{\rm C}\times$SU(2)$_{\rm L}\times$U(1)$_{\rm Y}$ are given by
\begin{equation}
\alpha _3 \simeq \frac{\alpha _{\rm GUT}}{1+\alpha _{\rm GUT}/\alpha _{3{\rm H}}},
~~~\alpha _2 = \alpha _{\rm GUT}, ~~~
\alpha _1 \simeq \frac{\alpha _{\rm GUT}}{1+\frac{1}{15}
\alpha _{\rm GUT}/\alpha _{1{\rm H}}},
\end{equation}
where $\alpha _{3{\rm H}} =g^2_{3{\rm H}}/4\pi$ and $\alpha _{1{\rm H}}
=g^2_{1{\rm H}}/4\pi$. We see that the GUT unification of the three gauge
coupling constants $\alpha _3, \alpha _2$ and $\alpha _1$ is
approximately realized in a strong coupling region of the hypercolor
gauge interactions, {\it i.e.} 
\raisebox{1pt}{$\alpha _{3{\rm H}}, \alpha _{1{\rm H}}$} $\gtrsim {\cal O}(1)$.

It is very important that massless fields of matter multiplets in the
above vacuum eq.(\ref{vacuum}) are only
a pair of hypercolor chiral multiplets, $Q^6_\alpha$ and $\bar{Q}^\alpha _6$,
which now transforms as a color-triplet under 
the unbroken SU(3)$_{\rm C}$. It is easy to give SUSY-invariant masses
to these massless triplets by introducing the following superpotential,
\begin{equation}
W = hQ^k_\alpha {\bar Q}^\alpha _6H_k + {\bar h}Q^6_\alpha 
{\bar Q}^\alpha _k{\bar H}^k.
\label{superpotential2}
\end{equation}
The color-triplets Higgs $H_a$ and ${\bar H}^a (a= 1,2,3)$ acquire masses of
order of the unification scale $M_{\rm GUT}\simeq V$ together with 
the sixth hyperquarks ${\bar Q}^\alpha _6$ and $Q^6_\alpha$ in the 
vacuum eq.(\ref{vacuum}). 
On the other hand, the weak-doublet
Higgs $H_i$ and ${\bar H}^i (i=4,5)$ remain massless since there are no
partners for them to form the SUSY-invariant masses. The Peccei-Quinn
like chiral symmetry or the R-symmetry may forbid the tree-level masses 
for Higgs multiplets 
$H_k$ and ${\bar H}^k$. The masslessness of the weak-doublets 
$H_i$ and ${\bar H}^i$ is
guaranteed by such symmetries as long as they are unbroken. 
We find that only a discrete ${\bf Z}_4$ R-symmetry is a consistent symmetry
to prevent the mass term for the Higgs multiplets.\footnote{We show in
section \ref{sec:gaugino-med} that the Higgs doublets acquire the 
SUSY-invariant mass of order of the gravitino mass after the 
${\bf Z}_{4 {\rm R}}$ symmetry is broken down to the ${\bf Z}_2$ R-parity.}
We show the ${\bf Z}_{4{\rm R}}$ charges for
the chiral multiplets in Table \ref{tab:R-charge}.
It is now clear that the superpotentials
eq.(\ref{superpotential}) and eq.(\ref{superpotential2}) are consistent with
this discrete ${\bf Z}_{4{\rm R}}$ symmetry.
(Notice that we have a continuous $\U(1)_{\rm R}$ in the limit of
$h=0$ \cite{iy,kurosawa}.)
We also easily find that the dimension-five 
operators for the proton decay \cite{syw} are forbidden by this
${\bf Z}_{4{\rm R}}$ 
symmetry. Furthermore, the superpotential
eq.(\ref{superpotential2}) renders the vacuum eq.(\ref{vacuum}) to be a unique 
one in the theory and hence there are no massless moduli fields 
besides the SUSY standard-model particles \cite{iy}.

We note that the above mechanism to give large masses to the
color-triplet Higgs multiplets keeping massless Higgs doublets is very
similar to the missing partner mechanism observed in the standard GUT
\cite{my}. However, there is a crucial difference. The masslessness of
the Higgs doublets is guaranteed by the R-symmetry in the present model, 
while there is no consistent symmetry suppressing the SUSY-invariant
Higgs mass in the missing partner model. 

We comment on phenomenological problems in the original $\SU(5)_{\rm GUT}
\times \U(3)_{\rm H}$ model. Since the
Higgs multiplets giving masses for quarks and leptons are ${\bf 5}$ and
${\bf 5^*}$ of the SU(5)$_{\rm GUT}$, we have the usual GUT relation, 
\begin{equation}
m_b = m_\tau,~~~m_s = m_\mu,~~~m_d = m_e,
\end{equation}
at the unification scale. This GUT relation is very successful for the
third family, $m_b = m_\tau$, but it must be largely violated for the
second and the first families. The possible lowest dimensional operators 
generating GUT-breaking effects in the quark- and lepton-mass matrices
are 
\begin{equation}
W = f_{ij}{\bf 5^*}_i{\bf 10}_j{\bar H}
\frac{\langle Q{\bar Q}\rangle}{M^2_*}.
\label{gut-breaking}
\end{equation}
Here, $M_*$ is the cut-off scale of the present theory. It is clear that 
if one takes $M_*\simeq M_{\rm Pl}\simeq 2.4\times 10^{18}$ GeV one obtains 
too small GUT-breaking effects as $|1-m_s/m_\mu| \lesssim 10^{-2}$. 
This already implies that the cut-off scale should be much lower than 
the Planck scale $M_{\rm Pl}$. In fact, to produce the observed mass 
spectrum for quarks and leptons we need \cite{kurosawa},
\begin{equation}
\frac {\langle Q{\bar Q}\rangle}{M_*^2} \gtrsim {\cal O}(10^{-2}),
\end{equation}
which leads to $M_*\lesssim 10^{17}$ GeV for $\langle Q{\bar Q}\rangle \simeq 
(10^{16} {\rm GeV})^2$. 

As mentioned in the introduction, this
lower-energy cut-off may solve another problem. That is, the GUT
unification of the three gauge couplings, $\alpha _3\simeq \alpha
_2\simeq \alpha_1$, to $5 \%$ accuracy requires  sufficiently large 
U(3)$_{\rm H}$ gauge coupling constants, with which the Landau pole of the
hypercolor U(1)$_{\rm H}$ gauge interactions appears at $\sim 6\times
10^{17}$ GeV \cite{luty}. Thus, the presence of the cut-off $M_*$ at 
$\sim 10^{17}$ GeV is obviously welcome to the present model.

We have shown, in this section, that somewhat mysterious, but
interesting features are required in the original $\SU(5)_{\rm GUT}
\times \U(3)_{\rm H}$ model for successful phenomenologies : the
disparity of gauge coupling constants of $\SU(5)_{\rm GUT}$ and
$\U(3)_{\rm H}$, the quasi-${\cal N}=2$ structure in the hypercolor
$\U(3)_{\rm H}$ sector, and the relatively small cut-off scale $M_*$
compared to the Planck scale $M_{\rm Pl}$. We consider that such
mysterious features are important indications of a more fundamental
theory. In the next section we show that they are all naturally explained by 
embedding the original model into the brane world in a higher
dimensional space-time.

\section{Embedding into the brane world}
\label{sec:embedding}
%
%
We embed the SUSY $\SU(5)_{\rm GUT}\times \U(3)_{\rm H}$ unification
model discussed in the previous section into the brane world in a higher 
dimensional space-time. Before
giving a detailed discussion we first determine the dimension $n$
of the extra space. 

Let us suppose that the fundamental theory is described in
$3+n$ dimensional space and a time. The Einstein-Hilbert action of gravity 
in $4+n$ dimensional space-time is given by
\begin{equation}
{\cal S} = M_*^{2+n}\int\int \sqrt{-g^{(4+n)}}{\cal R}d^4xd^ny,
\end{equation}
where $M_*$ is the gravitational scale in the $4+n$ dimensional
space-time, and $g^{(4+n)}_{\mu \nu}$ and ${\cal R}$ are the metric and 
the scalar curvature. We identify the gravitational scale in the $4+n$
dimensional space-time, $M_*$, with the fundamental cut-off scale discussed 
in the previous section, and we take $M_* \lesssim 10^{17}$ GeV in the
following discussion. The $y$ denotes coordinates of the extra space. 
The extra dimensions are assumed to be compactified with ${\cal V}=
\tilde{L}^n$ the volume of the extra space. We assume the metric in the extra dimension to be 
orthogonal to those for our 4-dimensional space-time:
\begin{equation}
d{s}^2 = g _{\mu \nu}(x)dx^\mu dx^\nu + dy^2,
\end{equation}
where $g(x)_{\mu \nu}$ is the metric in the 4-dimension. The 
integration over $dy$ leads to the action in 4-dimension,
\begin{equation}
{\cal S}_4 = M_*^{2+n}\tilde{L}^n\int\sqrt{-g}{\cal R}_4d^4x.
\end{equation}
The coefficient in front of the integral must be
the 4-dimensional Planck scale $M_{\rm Pl}$ so that 
\begin{equation}
M^2_{\rm Pl} = M_*^2(M_*\tilde{L})^n.
\label{planck}
\end{equation}
The Planck scale in the 4-dimensional space-time appears to be
an effective scale, rather than a fundamental one \cite{witten,nima}.

From eq.(\ref{planck}) we obtain
\begin{equation}
10^{\frac{2}{n}} \lesssim \left( \frac{M_{\rm Pl}}{M_*}\right)^{\frac{2}{n}}=
(M_*\tilde{L}) = (M_{\rm Pl}\tilde{L})^{\frac{2}{n+2}}\lesssim 10^{\frac{4}{n+2}}.
\label{size}
\end{equation}
Here, the lower bound comes from the condition $M_* \lesssim 10^{17}$
GeV and the upper bound is obtained from the condition $\tilde{L}^{-1} \lesssim
1/M_{\rm GUT}$.\footnote{If $\tilde{L}>1/M_{\rm GUT}$ we have Kaluza-Klein
towers of adjoint matter multiplets below the unification scale, since
we put later the $\SU(5)_{\rm GUT}$ vector multiplet in the bulk. They
receive large GUT-breaking effects in their mass spectrum and hence they
may affect significantly the gauge coupling unification.}
The above equation (\ref{size}) suggests $n\geq 2$. 
On the other hand, as the number of extra dimensions becomes larger, 
the effective compactification size $\tilde{L}$ becomes smaller (see
eq. (\ref{size})), which yields non-negligible contact interactions between
sfermions on our brane and some SUSY-breaking field on a hidden brane 
causing too large flavour-changing neutral currents (FCNC's) \cite{randall} 
as shown in the next section. Thus, we are led to consider the safest
case of $n=2$, that is a 6-dimensional 
space-time. Here, the size of the extra dimensions is determined as 
$M_*\tilde{L} \simeq 10$ and thus $\tilde{L}\simeq 1/M_{\rm
GUT}$ and $M_* \simeq 10^{17} \GeV$.
In the following discussion we concentrate ourselves on the 
6-dimensional space-time.\footnote{We will show, in section \ref{sec:string}, that our
choice of the 6-dimensional space-time is naturally extended in a 
string theory.} However, it may be straightforward to extend 
our analysis to different dimensional theories.

We discuss first the group theoretical structure of 6-dimensional SUSY
\cite{stra}.  
The 6-dimensional SUSY is specified by giving a pair of two integers
$({\cal N}_{{\bf 4}_+},{\cal N}_{{\bf 4}_-})$.
The integer ${\cal N}_{{\bf 4}_+({\bf 4}_-)}$ represents
the number of supercharges ${\cal Q}^{(6)}_{{\bf 4_+},A({\bf 4_-},B)}$ that 
belong to ${\bf 4_+(4_-)}$ spinor representation of the $\SO(5,1)$.
The index $A(B)$ runs over $1,2,...,2{\cal N}_{\bf 4_+(4_-)}$.
The pseudo-Majorana condition is imposed (see appendix A), and only half 
of them are independent:
\begin{eqnarray}
 {\cal Q}^{(6)}_{{\bf 4_+},A+{\cal N}_{\bf 4_+}}
      = - {\cal Q}^{(6)c}_{{\bf 4_+},A}
       \quad ({\rm for~} A = 1,2,\ldots,{\cal N}_{{\bf 4}_+}), 
\label{eq:symplectic-SUSY6-4}\\
 {\cal Q}^{(6)}_{{\bf 4_-},B+{\cal N}_{\bf 4_-}}
 = - {\cal Q}^{(6)c}_{{\bf 4_-},B}
      \quad ({\rm for~} B = 1,2,\ldots,{\cal N}_{{\bf 4}_-}) 
\label{eq:symplectic-SUSY6-4'}.
\end{eqnarray}
Here, the ${\cal Q}^{(6)c}$ on the right-hand sides denotes charge
conjugation of the ${\cal Q}^{(6)}$. For notations, see appendix A.
The R-symmetry of $({\cal N}_{{\bf 4}_+},{\cal N}_{{\bf 4}_-})$ SUSY is
$\Sp({\cal N}_{{\bf 4}_+})\times\Sp({\cal N}_{{\bf 4}_-})$ and 
SUSY charges ${\cal Q}^{(6)}_{{\bf 4_+},A({\bf 4_-},B)}$ belong to 
the fundamental representation of the $\Sp({\cal N}_{\bf 4_+(4_-)})$.

By a dimensional reduction of the $x_4$ and $x_5$ directions,
each supercharge is decomposed into two Weyl spinors in the following way:
\begin{eqnarray}
{\cal Q}^{(6)}_{{\bf 4_-},B}&\rightarrow& \left(\begin{array}{cc}
			 &   - {\cal Q}^{(4)B+{\cal N}_{\bf 4_-}}_{\alpha} \\
            - \bar{{\cal Q}}^{(4)\dot{\alpha}}_B &
				  \end{array}\right), \nonumber \\
{\cal Q}^{(6)}_{{\bf 4_-},B+{\cal N}_{\bf 4_-}}&\rightarrow& \left(\begin{array}{cc}
		&   {\cal Q}^{(4)B}_{\alpha} \\
          - \bar{{\cal Q}}^{(4)\dot{\alpha}}_{B+{\cal N}_{\bf 4_-}} &
				  \end{array}\right), 
  \nonumber \\
{\cal Q}^{(6)}_{{\bf 4_+},A}&\rightarrow& \left(\begin{array}{cc}
        {\cal Q}^{(4)A+2{\cal N}_{\bf 4_-}+{\cal N}_{\bf 4_+}}_{\alpha} & \\
           &  \bar{{\cal Q}}^{(4)\dot{\alpha}}_{A+2{\cal N}_{\bf 4_-}}
				  \end{array}\right), \nonumber \\
{\cal Q}^{(6)}_{{\bf 4_+},A+{\cal N}_{\bf 4_+}}&\rightarrow& 
   \left(\begin{array}{cc}
	 - {\cal Q}^{(4)A+2{\cal N}_{\bf 4_-}}_{\alpha} & \\
  & \bar{{\cal Q}}^{(4)\dot{\alpha}}_{A+2{\cal N}_{\bf 4_-}+{\cal N}_{\bf 4_+}}
				  \end{array}\right),
\label{susydecomp}
\end{eqnarray}
where ${\cal Q}^{(4)a}_{\alpha}(a=1,..., 2{\cal N}_{\bf 4_+}+
2{\cal N}_{\bf 4_-})$ are 4-dimensional SUSY charges that belong
to $({\bf 2,1})$ spinor representation of the $\SO(3,1)$. (
$\bar{{\cal Q}}^{(4)\dot{\alpha}}_a(a=1,..., 2{\cal N}_{\bf 4_+}+
2{\cal N}_{\bf 4_-})$ belong to $({\bf 1,2})$ of the $\SO(3,1)$.)
Thus, the number of independent supercharges in the 4-dimensional
space-time is ${\cal N}=2{\cal N}_{\bf 4_+}+2{\cal N}_{\bf 4_-}$.
After the dimensional reduction, the $\SO(2)_{45}$ rotation on the 
$x_4$-$x_5$ plane, which was a subgroup of the $\SO(5,1)$, now becomes an
internal symmetry in the 4-dimensional theory. Since SUSY charges
transform under the $\SO(5,1)$ as spinor representations, they are 
charged under the $\SO(2)_{45}$.  Namely, ${\cal Q}^{(4)a}_{\alpha}(a=1,..., 
2{\cal N}_{\bf 4_-})$ carry the $\SO(2)_{45}$ charges $-1$, while 
${\cal Q}^{(4)b+ 2{\cal N}_{\bf 4_-}}_{\alpha}(b=1,..., 2{\cal N}_{\bf
4_+})$ the charges +1.

We now consider the $\SU(5)_{\rm GUT}$ gauge vector multiplet that is
supposed to live in the 6-dimensional bulk. 
We assume that the 6-dimensional space-time has an ${\cal N}=4$ SUSY 
in the 4-dimensional sense before the compactification/orbifolding of 
the extra dimensional space.
We will see later that such a higher ${\cal N}$ SUSY is necessary for 
having a quasi-${\cal N}=2$ structure in the hypercolor $\U(3)_{\rm H}$ 
sector. 
There are two choices, $(1,1)$ and $(2,0)$, for the
SUSY in 6-dimension. 
If the extra 2-dimensional space is compactified,
these two choices become the same ${\cal N}=4$ SUSY.
In the case of Abelian gauge theories, it is clear that 
the massless spectra after the compactification are identical to each other.
Although we do not know the definition of the $(2,0)$ theory
for the non-Abelian case,
it is expected that the $(2,0)$ theory and the $(1,1)$ theory
are equivalent to each other via some duality
similar to the T-duality in string theories \cite{20-11-dual}, 
when the extra 2-dimensional space is compactified.
Therefore, we take the $(1,1)$-SUSY theory in what follows.
The R-symmetry of this 6-dimensional theory is
$\Sp(1)_{\bf 4_+}\times\Sp(1)_{\bf 4_-}
\simeq \SU(2)_{\bf 4_+}\times\SU(2)_{\bf 4_-} \simeq \SO(4)_{\rm R}$.

The vector multiplet is a unique supermultiplet of the $(1,1)$ SUSY 
theory besides the gravity multiplet.
It consists of a 6-dimensional vector field $A_\mu$ ($\mu=0,...,5$),
two pseudo-Majorana-Weyl spinors with opposite chirality
$\Lambda_{{\bf 4_+},B}$ and $\Lambda_{{\bf 4_-},A}$ ($A,B=1,2$)\footnote{
This means that the Weyl spinors, $\Lambda_{{\bf 4_+},B}$ and 
$\Lambda_{{\bf 4_-},A}$ ($A,B=1,2$),
satisfy similar relations to eq.(\ref{eq:symplectic-SUSY6-4}) and 
eq.(\ref{eq:symplectic-SUSY6-4'}), respectively.}
and two complex scalar fields $\sigma'$ and $\sigma''$.
All are certainly adjoint representations of the $\SU(5)_{\rm GUT}$.
 The vector field $A_\mu$ is an
invariant under the R-symmetry $\SU(2)_{\bf 4_+} \times \SU(2)_{\bf 4_-}$.
The spinor $\Lambda_{{\bf 4_-},A}$ transforms as a doublet of 
the $\SU(2)_{\bf 4_+}$ and the other spinor $\Lambda_{{\bf 4_+},B}$ as a
doublet of the $\SU(2)_{\bf 4_-}$. Four real components of scaler fields
belong to ({\bf 2},{\bf 2}) representation of the R-symmetry, which 
transforms as
\begin{equation}
\left(\begin{array}{cc}
 \sigma' & \sigma'' \\  \sigma^{''*} & - \sigma^{'*}
      \end{array}
\right)_{BA} 
\rightarrow 
\SU(2)_{{\bf 4_-},B}^{~~~~~B'} 
\left(\begin{array}{cc}
 \sigma' & \sigma'' \\  \sigma^{''*} & - \sigma^{'*}
      \end{array}
\right)_{B'A'}
\SU(2)_{{\bf 4}_+,A}^{~~~~~A'}. 
\end{equation}
In other words, they form a vector of the $\SO(4)_{\rm R} \simeq
\SU(2)_{\bf 4_+}\times \SU(2)_{\bf 4_-}$.

If the extra 2-dimensional space is compactified on a torus ${\bf T}^2$, 
we have 
an ${\cal N}=4$ SUSY theory in the 4-dimensional space-time; the flat metric
of torus preserves all the SUSY. This 4-dimensional theory contains an
${\cal N}=4$ gauge vector multiplet consisting of a 4-dimensional 
vector field $A_i (i = 0,...,3)$, four fermion partners 
$\chi_{\alpha,a}$ ($a=1,...,4$) and three complex scalar 
fields $\sigma ,\sigma', \sigma''$.
The four Weyl fermions $\chi_{\alpha,a}$ are obtained from the
6-dimensional fermion fields $\Lambda_{\bf 4_+(4_-)}$ by a decomposition 
similar to eq.(\ref{susydecomp}) as 
\begin{equation}
\Lambda_{{\bf 4_+},1} \rightarrow 
\left( \begin{array}{cc}
 -i \chi_{\alpha,1} & \\ 
 & -i \bar{\chi}^{\dot{\alpha},2}
	\end{array}\right)  \quad 
\Lambda_{{\bf 4_-},1} \rightarrow
\left(\begin{array}{cc}
 & -i \chi_{\alpha,3}  \\
 -i \bar{\chi}^{\dot{\alpha},4}
      \end{array}\right), 
\end{equation}
and the complex $\sigma$ field is
constructed by the original 6-dimensional vector field as 
\begin{equation}
\sigma(x,y)\equiv \frac{1}{\sqrt{2}}(A_4(x,y) +iA_5(x,y)).
\end{equation}
In terms of ${\cal N}=1$ SUSY multiplets, the ${\cal N}=4$ vector 
multiplet consists of one vector multiplet ${\cal W}_\alpha=$
$(i\chi_{1},F_{ij})$ and three chiral multiplets 
$\Sigma =(i\sigma,\chi_2)$, $\Sigma'=(i\sigma',\chi_3)$ and  
$\Sigma''=(i\sigma'',\chi_4)$.
Thus, we have three adjoint chiral multiplets. If they exist as 
massless particles,
they change the renormalization-group-equation (RGE) flow of the 
standard-model gauge coupling constants, 
leading to a blow up of gauge couplings well below the unification scale
$M_{\rm GUT}$. This too many adjoint multiplets is a direct consequence 
of the ${\cal N}=4$ SUSY in 4-dimension, and therefore, we do not want the 
${\cal N}=4$ SUSY to be preserved below the compactification scale
$\tilde{L}^{-1}$($\sim $ the unification scale $M_{\rm GUT}$).

To obtain the ${\cal N}=1$ SUSY theory 
we have to do some orbifolding \cite{orbifold} by using a discrete
subgroup of the symmetry $\SO(2)_{45} \times \SU(2)_{\bf 4_+} \times 
\SU(2)_{\bf 4_-}$.\footnote{This orbifolding is also necessary to obtain 
chiral representations ${\bf 5}^* + {\bf 10}$ in SUSY standard-model sector.}
Without specifying a concrete form of the orbifold group,
we can show that only the ${\cal N}=1$ vector multiplet ${\cal W}_\alpha$ 
remains massless after the orbifolding as follows. 
SUSY charges and/or bulk fields that are invariants of the orbifold-group
action survive the orbifolding. That is, SUSY charges that are charged
under the action disappear at low energies and bulk particles charged 
under the action do not have massless modes in their Kaluza-Klein spectra, 
as we see later.
Suppose that we have an orbifolding after which we have only one SUSY charge 
left unbroken among four. Then, only one fermion is left massless after the orbifolding,
because the fermion fields $\bar{\chi}^a$ transform under the 
the orbifolding symmetry in the identical way to the supercharges ${\cal Q}^a$.
This massless fermion must belong to the vector multiplet ${\cal
W}_\alpha =(i\chi_1,F_{ij})$
because the vector field $A_i$ is invariant under the symmetry.
On the other hand, all scalar fields $\sigma$, $\sigma'$ and $\sigma''$ are
projected out along with the three fermion partners of the unbroken
${\cal N}=1$ SUSY.
Therefore, if we keep the ${\cal N}=1$ SUSY after the orbifolding,
then only the vector multiplet ${\cal W_\alpha}$ remains massless 
and we do not have to worry about the emergence of unwanted adjoint 
chiral multiplets.

Let us see how the above procedure works, taking an explicit example 
of the orbifolding. The orbifolding is, in terms of the 6-dimensional
theory, given by gauging a symmetry which is a combination of space 
($x_4$-$x_5$) rotation and reflection ({\it i.e.} $\O(2)_{45}$) and 
internal symmetry transformations. After the compactification 
(dimensional reduction),
however, the space symmetry of the compactified 2-dimensional
space 
is treated as an internal symmetry, or more specifically, as an
$\O(2)_{45}$ R-symmetry.
Thus, the orbifold group is considered as a subgroup of purely internal
symmetries. We will seek for a suitable orbifold group using the 4-dimensional
terms, because of the convenience.

We now search for a candidate for the orbifold group which is a subgroup 
of the symmetry group (space-rotation symmetry $\SO(2)_{45}$) $\times$ 
(6-dimensional R-symmetry $\SU(2)_{\bf 4_+} \times \SU(2)_{\bf 4_-}$). 
This group is embedded into the maximally 
possible R-symmetry $\SU(4)_{\rm R}$ of 
the 4-dimensional ${\cal N}=4$ SUSY.  In the following analysis, we
consider the whole $\SU(4)_{\rm R}$ group.\footnote{
There are two reasons why we proceed our discussion in terms of
$\SU(4)_{\rm R}$. (1) The space symmetry $\SO(2)_{45}$ can be treated 
just as an internal R-symmetry. (2) Relations between the 
6-dimensional R-symmetry $\SU(2)_{\bf 4_+} \times \SU(2)_{\bf 4_-}$ 
and the R-symmetry $\U(1)_{\rm R} \times \SU(2)_{\rm R} \times 
\SU(2)_{\rm F}$ in the hypercolor $\U(3)_{\rm H}$
sector, which appears later, can be explicitly described in terms of 
$\SU(4)_{\rm R}$. Discussion in section \ref{sec:string} gives 
a geometrical explanation of the relations among various R-symmetries.} 
Notice that as long as we take the orbifold group within
the maximal torus of the $\SU(4)_{\rm R}$, which is the case for 
the orbifolding 
we adopt in this paper, the orbifold group is always contained in
the original $\SO(2)_{45} \times \SU(2)_{\bf 4_+} \times \SU(2)_{\bf 4_-}$, 
since the maximal torus of the $\SU(4)_{\rm R}$ 
is the same as that of the original group.\footnote{
In fact, it is not always necessary that the orbifold group be contained in the
maximal torus of the $\SO(2)_{45} \times \SU(2)_{\bf 4_+} \times
\SU(2)_{\bf 4_-}$.} 
The SUSY charges ${\cal Q}^{(4)a}$ transform as ${\bf 4}$ 
under the $\SU(4)_{\rm R}$, the fermions $\chi_a$ as ${\bf 4}^{*}$ and the scalars 
\begin{equation}
 \sigma_{ab} \equiv \left(\begin{array}{cccc}
		0 & \sigma & \sigma' & \sigma'' \\
                -\sigma & 0 & \sigma^{''*} & -\sigma^{'*} \\
                -\sigma' & -\sigma^{''*} & 0 & \sigma^* \\
                -\sigma'' & \sigma^{'*} & -\sigma^* & 0 \\
		     \end{array}\right)
\label{eq:N4-scalar-multiplet}
\end{equation}
as the 2nd rank anti-symmetric tensor of ${\bf 4}^*$ $({\it i.e. }
\wedge^2 {\bf 4}^{*})$. Here, the $\SO(2)_{45}\times 
\SU(2)_{\bf 4_+} \times \SU(2)_{\bf 4_-}$ are embedded in the 
$\SU(4)_{\rm R}$ as
\begin{equation}
\left(\begin{array}{cccc}
 e^{-i\frac{\varphi}{2}}& & & \\
  & e^{-i\frac{\varphi}{2}}& & \\
  & & e^{i\frac{\varphi}{2}}& \\
  & & & e^{i\frac{\varphi}{2}}
      \end{array}\right),
\left( \begin{array}{cc}
 \SU(2)_{\bf 4_-}& \\ & \SU(2)_{\bf 4_+}
       \end{array}\right) \subset 
\left(\SU(4)_{\rm R}\right),
\end{equation}
where $0 \leq \varphi < 4\pi $.

Let us consider an $\SU(3)_{\rm R}$ subgroup of the $\SU(4)_{\rm R}$, 
\begin{equation}
\left(\begin{array}{cc}
        1 &  \\
          & \SU(3)_{\rm R}  
       \end{array} \right) 
\subset \left( \SU(4)_{\rm R} \right).
\end{equation}
If the orbifold group is contained in this $\SU(3)_{\rm R}$ subgroup, then
the SUSY charge ${\cal Q}^{(4)1}$ is invariant under the orbifold group, and
hence the ${\cal N}=1$ SUSY survives the orbifolding. 
We take a ${\bf Z}_4$ subgroup of the $\SU(3)_{\rm R}$ as the orbifold 
group\footnote{This ${\bf Z}_4$ group is taken just as
an example. There are many other examples of orbifold group that
satisfy phenomenological requirements.} whose generator is  
\begin{equation}
 \left( \begin{array}{cccc}
        1 & & & \\
          & i & & \\
          & & i & \\
          & & & -1
       \end{array} \right) \in \left( \SU(4)_{\rm R}\right).
\label{Z4-generator}
\end{equation} 
This ${\bf Z}_4$ subgroup is in the maximal torus of the $\SU(3)_{\rm
R}$, and hence  it is contained in the $\SO(2)_{45}\times \SU(2)_{\bf 4_+}\times 
\SU(2)_{\bf 4_-}$.
This generator is decomposed into a product of $\SO(2)_{45}$ and 
$\SU(2)_{\bf 4_+}\times \SU(2)_{\bf 4_-}$ elements as: 
\begin{equation}
 \left( \begin{array}{cccc}
        1 & & & \\
          & i & & \\
          & & i & \\
          & & & -1
       \end{array} \right) = 
 \left( \begin{array}{cccc}
        e^{i\frac{\pi}{4}}  & & & \\
          & e^{i\frac{\pi}{4}} & & \\
          & & e^{-i\frac{\pi}{4}} & \\
          & & & e^{-i\frac{\pi}{4}}
       \end{array} \right)
 \left( \begin{array}{cccc}
        e^{-i\frac{\pi}{4}} & & & \\
          & e^{i\frac{\pi}{4}} & & \\
          & & e^{ i\frac{3\pi}{4}} & \\
          & & & e^{-i\frac{3\pi}{4}}
       \end{array} \right). 
\end{equation}
Therefore, the above orbifolding is equivalent to the identification of 
extra-dimensional space ${\bf T}^2$ under the $-\pi/2 $ rotation\footnote{
Note that the $\sigma \propto (A_4 + iA_5)$ transforms as 
$\sigma \rightarrow e^{i\varphi} \sigma$ under the $\SO(2)_{45}$ 
when the $\SO(2)_{45}$ element is represented as diag($e^{-i\varphi/2},e^{-i\varphi/2},
e^{i\varphi/2},e^{i\varphi/2}$) on the ${\bf 4}$ representation of the 
$\SU(4)_{{\rm R}}$.}  in the
$x_4$-$x_5$ plane accompanied by a suitable twist under the internal 
$\SU(2)_{\bf 4_+}\times \SU(2)_{\bf 4_-}$ transformation.
The $\SU(3)_{\rm R}$ action on the fermions and the scalars are given by
\begin{equation}
\left(\begin{array}{c}
	\chi_1 \\ \chi_2 \\ \chi_3 \\\chi_4
				\end{array}\right)
\rightarrow
\left( \begin{array}{cc}
  1 & \\ & \SU(3)_{\rm R}^* 
	\end{array}\right)\left(\begin{array}{c}
	\chi_1 \\ \chi_2 \\ \chi_3 \\\chi_4
				\end{array}\right),
\end{equation}
\begin{equation}
\left(\begin{array}{c}
  \sigma \\ \sigma' \\ \sigma'' \\ \sigma^* \\ \sigma^{'*} \\ \sigma^{''*}
      \end{array}\right)
\rightarrow
 \left(\begin{array}{cc}
  \SU(3)_{\rm R}^* & \\ 
            & \SU(3)_{\rm R}
       \end{array}\right)
\left(\begin{array}{c}
  \sigma \\ \sigma' \\ \sigma'' \\ \sigma^* \\ \sigma^{'*} \\ \sigma^{''*}
      \end{array}\right).
\label{eq:so(6)-vector}
\end{equation}

Now, it is clear that we can eliminate the unwanted $\sigma$ and $\chi$ 
bulk fields from massless spectrum by the ${\bf Z}_4$ orbifolding of the torus
${\bf T}^2$. 
This is because wave functions of the ${\bf Z}_4$ charged
particles in the bulk must vanish at the orbifold fixed points:
Zero-modes in their Kaluza-Klein towers are eliminated and their 
lowest-energy states have Kaluza-Klein masses of order the 
$\tilde{L}^{-1}\simeq M_{\rm GUT}$ disappearing from the low-energy spectrum. 
Therefore, we have an ${\cal N}=1$ SUSY
theory in the bulk with only an $\SU(5)_{\rm GUT}$ gauge vector multiplet
below the Kaluza-Klein mass scale $\tilde{L}^{-1}\simeq M_{\rm GUT}$.

\begin{figure}[htb]
\centerline{\epsfbox{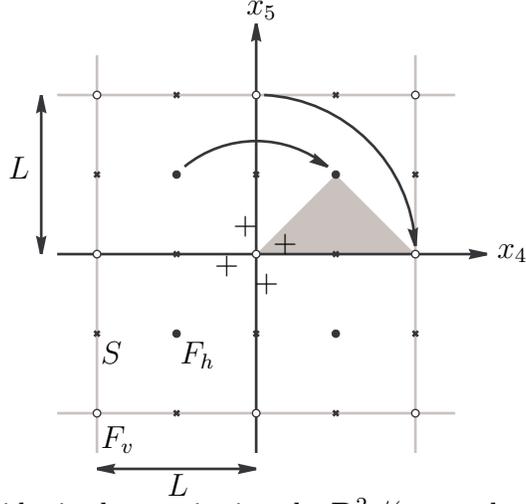}%
\put(-5,90){$x_4$}
\put(-100,182){$x_5$}
\put(-90,92){$+$}\put(-105,99){$+$}\put(-112,84){$+$}\put(-97,77){$+$}
\put(-125,50){$F_h$}
\put(-155,18){$F_v$}
\put(-155,50){$S$}
\put(-130,0){$L$}
\put(-190,120){$L$}
}
\caption{${\bf T}^2$ torus we consider in the text is given by 
${\bf R}^2$ /(square lattice) shown in this figure. ${\bf Z}_4$ 
orbifold group action on the $x_4$-$x_5$ plane and the fundamental
region of this ${\bf Z}_4$ are also described. There are three distinct 
singularities among which $F_v$ and $F_h$ are ${\bf Z}_4$ fixed points. 
Our visible sector 3-brane is located at the ${\bf Z}_4$ fixed point 
$F_v$ and the hidden sector 3-brane at the other ${\bf Z}_4$ fixed point 
$F_h$. $S$ is another singularity in the orbifold which is fixed under
 ${\bf Z}_2 \subset {\bf Z}_4$ but not fixed under the whole ${\bf Z}_4$. 
+'s are the location of the hypercolor $\U(3)_{\rm H}$ 3-brane 
and its mirror images under the ${\bf Z}_4$ transformation (their images 
 under  ${\bf Z} \times {\bf Z}$ lattice translation are omitted).
\label{Fig:T2-Z4-vis-U3}}
\end{figure}

We adopt, in most of this paper, the above ${\bf Z}_4$ orbifolding. 
This means that the torus ${\bf T}^2$ in the $x_4$-$x_5$ space
directions must have the ${\bf Z}_4$ rotational symmetry. 
Therefore, the ${\bf T}^2$ torus is the quotient of ${\bf R}^2$ by 
the square lattice, as shown in Fig.\ref{Fig:T2-Z4-vis-U3}.
The ${\bf T}^2/{\bf Z}_4$ orbifold has two distinct ${\bf Z}_4$ 
fixed points $F_v$ and $F_h$.\footnote{To be precise, ${\bf T}^2/{\bf
Z}_4$ orbifold has another singularity, which is not fixed under the
${\bf Z}_4$ but fixed under the ${\bf Z}_2 \subset {\bf Z}_4$.} 
We put a 3-brane at one of the fixed points $F_v$ and
assume that the chiral multiplets of quarks, leptons and a pair of 
Higgs $H_k$ and ${\bar H}^k (k=1,...,5)$ all reside on this 3-brane. 
The low-energy theory on the 3-brane is described by the 4-dimensional 
${\cal N}=1$ SUSY theory, where the SUSY is generated by one of the 
four SUSY charges, namely the ${\cal Q}^{(4)1}$, since the other three 
SUSY charges are not invariant under the ${\bf Z}_4$ orbifold group 
transformation.

We now show that the $\SU(5)_{\rm GUT}$ gauge coupling $g_{\rm GUT}$ 
in the 4-dimensional space-time receives a volume suppression. The gauge
coupling in the fundamental theory is defined as
\begin{equation}
{\cal S} = M_*^2\int d^4 x \int _{{\bf T}^2/{\bf Z}_4} d^2 y \int d^2\theta \sqrt{-g^{(6)}}
\frac {1}{g_0^2}{\cal W}_\alpha {\cal W}^\alpha ,
\end{equation}
where ${\cal W}_\alpha$ and $g_0$ are the field strength of the vector
multiplet and the dimensionless gauge coupling of the fundamental
theory, respectively. The integration of $d^2y$ yields a 4-dimensional action as
\begin{equation}
{\cal S}_4 = \frac{1}{4} (M_*L)^2 \int d^4 x \int d^2\theta \sqrt{-g}
\frac {1}{g_0^2}{\cal W}_\alpha {\cal W}^\alpha ,
\end{equation}
where $L$ is the lattice spacing and $1/{4}$ the volume of
${\bf T}^2/{\bf Z}_4$ measured in the unit of $L^2$ ({\it i.e.} ${\cal
V}=\tilde{L}^2=L^2/4$).
Thus, the 4-dimensional gauge coupling defined as $\alpha _{\rm GUT}
\equiv 4 \alpha _0(M_*L)^{-2}$  has a suppression factor 
$4/(M_*L)^2 \simeq 10^{-2}$. The
experimental value $\alpha _{\rm GUT}\simeq 1/24$ gives the gauge
coupling $\alpha _0 \equiv g_0^2/4\pi \simeq 4$. This suggests that the
fundamental theory is strong entirely, but as shown in \cite{kaplan} it
may be still weak enough to perform a perturbative analysis. Above the
compactification scale ({\it i.e.} the unification scale $M_{\rm GUT}$),
Kaluza-Klein towers including excitations of the adjoint chiral
multiplets $\Sigma$, $\Sigma'$ and $\Sigma''$ contribute in general 
to RGE's for the 4-dimensional gauge coupling 
$\alpha_{\rm GUT} (\mu )$,
where $\mu$ is the renormalization energy scale \cite{dienes}. However,
the Kaluza-Klein towers form ${\cal N}=4$ SUSY multiplets and their
contributions to the RGE's are canceled out
completely and vanish at all orders of the perturbation theory. Thus, we
have the usual RGE's in the 4-dimensional
space-time and the coupling runs logarithmically. 
The difference between $\alpha_{\rm GUT}(\mu =M_*)$ and $\alpha_{\rm GUT}(\mu
=M_{\rm GUT})$ is quite small, and $\alpha_0$ at the
fundamental scale $M_* \simeq 10^{17}\GeV$ is about the same as that at
the unification scale $\mu = M_{\rm GUT}$.

Now, we discuss SUSY in the hypercolor $\U(3)_{\rm H}$ sector which
is supposed to reside on a 3-brane in the 6-dimensional bulk 
(see Fig.\ref{Fig:T2-Z4-vis-U3}). 
We have stated previously that we need an ${\cal N}=4$ SUSY in the bulk
theory (after a reduction to the 4 dimensions) for having naturally  
the quasi-${\cal N}=2$ SUSY structure in the hypercolor sector.
Indeed, the $\U(3)_{\rm H}$ 3-brane has tension, and because of 
this energy density,  
not all of (at most half of) the SUSY charges in the bulk theory 
is left unbroken on the $\U(3)_{\rm H}$-sector fields (see appendix B 
for details). Therefore, $(0,1)$ SUSY in the bulk, which would become 
the 4-dimensional ${\cal N}=2$ SUSY after the dimensional reduction, 
is not enough for our purpose. This is the reason why we need one more 
SUSY in the original 6-dimensional bulk theory. We assume half of the 
${\cal N}=4$ ({\it i.e.} ${\cal N}=2$) SUSY to be realized on 
the hypercolor 3-brane, taking the $(1,1)$ SUSY in the 6-dimensional
bulk. We show in section \ref{sec:string} that this is indeed the case 
in string theories.

Although the orbifolding reduces the ${\cal N}=4$ SUSY in the bulk 
theory to the ${\cal N}=1$ SUSY in the effective 4 dimensional theory, 
the SUSY realized on the 3-brane alone is not affected by the
orbifolding. This is because the $\U(3)_{\rm H}$-sector fields are 
confined on the 3-brane and do not see the global structure of 
the $x_4$-$x_5$ space but see only the local neighbourhood around 
the 3-brane. The discrete action of the orbifold group relates 
the 3-brane to its mirror images. We can choose wave functions on 
those mirror-image 3-branes so that the local structures around those 
3-branes are all equivalent to each other under the orbifold-group action.  
Thus, we see that there is no effect of the orbifolding as long as 
we are concerned only in the physics on the 3-brane ({\it i.e.} physics 
of the $\U(3)_{\rm H}$ vector multiplets and $Q^i,\bar{Q}_i, Q^6,\bar{Q}_6$ 
chiral multiplets).

The breaking of $P_4$ and $P_5$ gives rise to Nambu-Goldstone (NG) bosons,
which correspond to fluctuation modes of the $\U(3)_{\rm H}$ 3-brane.
They are always massless if the parallel transport symmetry is
broken only by the presence of the 3-brane.
However, the symmetry is already broken
by the orbifolding of the torus.
Therefore, we can expect that the NG bosons get masses and the position of
the brane is stabilized by some yet unknown dynamics (see also the
discussion in section \ref{sec:string}). We assume
throughout this paper that it is indeed the case.\footnote{Otherwise,
we have massless $\U(3)_{\rm H}$ adjoint chiral multiplets, which are
charged under the $\SU(3)_{\rm C}$ but not under the $\SU(2)_{\rm L} \times
\U(1)_{\rm Y}$, ruining the approximate unification of the 
standard-model gauge coupling constants.} 

The field theory on the $\U(3)_{\rm H}$ 3-brane alone is, therefore, 
described by an ${\cal N}=2$ SUSY theory in the 4-dimensional space-time. 
Here, we have ${\cal N}=2$ vector multiplets of the $\U(3)_{\rm H}$ 
gauge theory and six hypermultiplets 
($Q_\alpha^\rho$,${\bar Q}_\rho^\alpha$) ($\alpha =1,2,3; \rho
=1,...,6$). The ${\cal N}=2$ vector multiplets are decomposed into an 
${\cal N}=1$ vector multiplet ${\cal W}^{\SU(3)_{\rm H}}_\alpha$ of the 
$\SU(3)_{\rm H}$ and a vector multiplet ${\cal W}^{\U(1)_{\rm H}}_\alpha$ 
of the $\U(1)_{\rm H}$, and ${\cal N}=1$ chiral multiplets
$X^\alpha_\beta$ and $X_0$. We have a superpotential eq.(\ref{superpotential}) 
in the previous section with the ${\cal N}=2$ SUSY coupling relation:
\begin{equation}
\lambda =\lambda '=\sqrt{2}g_{3{\rm H}},~~~
\kappa =\kappa '=\sqrt{2}g_{1{\rm H}}.
\label{eq:n=2relation}
\end{equation}
The ${\cal N}=2$ SUSY on this brane is, however, broken by gauge
interactions of the bulk $\SU(5)_{\rm GUT}$ since the bulk SUSY is only  
${\cal N}=1$. Thus, the above Yukawa coupling constants, $\lambda,
\lambda '$ and $\kappa, \kappa '$, receive renormalization effects
differently from the $\SU(5)_{\rm GUT}$ gauge-multiplet loops. And the ${\cal
N}=2$ SUSY relation eq.(\ref{eq:n=2relation}) survives no longer at 
low energies. However, it is
extremely interesting that the presence of the $\U(3)_{\rm H}$ adjoint
chiral multiplets $X^\alpha _\beta$ and $X_0$ is an automatic consequence 
of the theory.
Recall that we have introduced them, by hand, to cause the
desired symmetry breaking in the original model.

We turn to discuss R-symmetry on the hypercolor 3-brane alone.
The possible maximum R-symmetry of the 4-dimensional ${\cal N}=2$ SUSY is  
$\U(1)_{\rm R} \times \SU(2)_{\rm R} \times \SU(2)_{\rm F} $, which is a
subgroup of the $\SU(4)_{\rm R}$ as 
\begin{equation}
\left( \begin{array}{cccc}
 e^{i\varphi/2}& & & \\
  & e^{-i\varphi/2}& & \\
  & & e^{-i\varphi/2}& \\
  & & & e^{i\varphi/2} 
       \end{array}\right),
\left(\begin{array}{ccc}
 * & & * \\
   & \SU(2)_{\rm F} & \\
 ** & & *
      \end{array}\right) \subset \left( \SU(4)_{\rm R} \right),
\label{eq:embed-U(1)89*SU(2)R*SU(2)L}
\end{equation}
where $0 \leq \varphi < 4 \pi$, and the $*$'s in the four corners of
the $4\times 4$ matrix represent the $\SU(2)_{\rm R}$.
As shown in the appendix B, one SUSY charge from ${\cal Q}^{(4)1}$ and 
${\cal Q}^{(4)2}$ and one from ${\cal Q}^{(4)3}$ and ${\cal Q}^{(4)4}$ 
form the ${\cal N}=2$ SUSY, and in the above choice of the embedding of 
the $\U(1)_{\rm R} \times \SU(2)_{\rm R} \times \SU(2)_{\rm F}$ into 
the $\SU(4)_{\rm R}$, they are ${\cal Q}^{(4)1}$ and ${\cal Q}^{(4)4}$.
Notice that the $\SU(2)_{\rm F}$ is not R-symmetry acting on the 
relevant ${\cal N}=2$ SUSY charges ${\cal Q}^{(4)1}$ and ${\cal
Q}^{(4)4}$, but a flavour symmetry that acts only on the $\SU(5)_{\rm
GUT}$ adjoint chiral multiplets $\Sigma$ and $\Sigma'$. 
The $\SU(2)_{\rm R}$ factor is the ordinary $\SU(2)$ R-symmetry in 
${\cal N}=2$ SUSY theories that 
acts on the ${\cal Q}^{(4)1}$ and ${\cal Q}^{(4)4}$. This symmetry is 
broken down to a $\U(1)$ symmetry by the FI F-term 
(see eq.(\ref{superpotential})). This $\U(1)$ symmetry mixes scalars 
of chiral multiplet $Q^k_\alpha$ ($\bar{Q}_k^\alpha$) and scalars of 
anti-chiral multiplets $\bar{Q}^{\alpha\dagger}_k$
($Q^{k\dagger}_\alpha$), respectively, and therefore, this symmetry has 
nothing to do with the usual low-energy R-symmetry.
Thus, we regard the $\U(1)_{\rm R}$ in 
eq.(\ref{eq:embed-U(1)89*SU(2)R*SU(2)L}) as the origin of the ${\bf
Z}_4$ R-symmetry discussed in the previous section. Indeed, R-charges 
are determined from the symmetry algebra: the $\U(3)_{\rm H}$ adjoint 
chiral multiplets $X_\alpha^\beta$ and $X_0$ have the $\U(1)_{\rm R}$-charges 
2, $(Q^i\bar{Q_i})$ and $(Q^6\bar{Q}_6)$ carry the $\U(1)_{\rm
R}$-charges 0.\footnote{
Embedding of the $\U(1)_{\rm R}$ into the $\SU(4)_{\rm R}$ 
determines also the the $\U(1)_{\rm R}$ charges of the $\SU(5)_{\rm
GUT}$ adjoint chiral multiplets: $\Sigma''$ has the charge 2, while 
$\Sigma$ and $\Sigma'$ have the charges 0.}
This charge assignment is the same as that given in 
Table \ref{tab:R-charge} for the $\U(3)_{\rm H}$-sector 
particles.\footnote{This coincidence of the charge assignment 
is not accidental, but rather a consequence of the ${\cal N}=2$ 
SUSY of the theory.} 
We do not find any reason that the $\U(1)_{\rm R}$ should be broken down 
to its discrete ${\bf Z}_{4{\rm R}}$ subgroup at this level. However, 
it becomes clear in section \ref{sec:string} that this $\U(1)_{\rm R}$ is in fact
broken down to a discrete ${\bf Z}_{4{\rm R}}$ in a natural extension 
of our 6-dimensional brane-world model to a 10-dimensional string theory.

The final comment in this section is that the hypercolor $\U(3)_{\rm H}$ 
3-brane should be very close to our 3-brane at an orbifold fixed point. 
This is because the nonrenormalizable operators in
eq.(\ref{gut-breaking}) are exponentially suppressed,
otherwise. However, as discussed in section \ref{sec:5-3}, these operators are 
necessary to produce the realistic mass spectra for quarks and
leptons. In this paper we assume the distance $\tilde{D}$  between the
hypercolor $\U(3)_{\rm H}$ brane and our brane to be $\tilde{D} \lesssim
1/M_*$ so that the exponential factor $\exp(-M_*\tilde{D})$ is 
of  ${\cal O}(1)$. 
This arrangement of closely separated two 3-branes is also crucial to
have an unsuppressed superpotential eq.(\ref{superpotential2}) between
the Higgs and the hyperquark chiral multiplets.

\section{Gaugino mediation of the SUSY breaking}
\label{sec:gaugino-med}
%
%
We discuss, in this section, the mediation  of SUSY breaking that is
supposed to occur on a 3-brane at the fixed point $F_h$ of our 
orbifold ${\bf T}^2/{\bf Z}_4$ (see Fig.\ref{Fig:T2-Z4-vis-U3}). We show 
that the present model is regarded as an extension of the
gaugino-mediation model of the SUSY breaking \cite{kaplan,kawasaki}.  

In supergravity, the hidden sector responsible for 
the SUSY breaking is assumed to be fully separated from the visible 
sector in order to suppress the unwanted FCNC's \cite{niels}. This 
separation is beautifully realized in the brane world as pointed out in 
ref.~\cite{randall}. We put the SUSY-breaking sector on a hidden
3-brane at the orbifold fixed point separated from 
the fixed-point 3-brane on which our visible sector resides. 
Then, it is obvious to realize the separation of the hidden and 
visible sectors, if the distance $D$ between the two 3-branes at the 
orbifold fixed points is sufficiently large and interactions between 
the two sectors are exponentially suppressed \cite{randall}. It seems also 
natural to postulate that the separation takes place in the
``conformal'' frame in supergravity \cite{randall,luty-sundrum}. 
If it is the case, the \kahler  potential ${\cal K}$ and superpotential
$W$ have the following forms \cite{kawasaki}:
\begin{eqnarray}
  {\cal L} = \int d^2\Theta\, 2{\cal E}
    \left[ -\frac{1}{8}(\bar{\cal D}\bar{\cal D}-8R)\,
           {\cal K}(\Phi , \Phi ^{\dagger}, Z, Z^{\dagger})
           + W(\Phi , Z) \right] + {\rm h.c.};
\end{eqnarray}
\begin{eqnarray}
  {\cal K}(\Phi , \Phi^{\dagger}, Z, Z^{\dagger})
     &=& -3 + f_{\rm vis}(\Phi , \Phi ^{\dagger}) + 
              f_{\rm hidd}(Z, Z^{\dagger}),
\label{kaler} \\
  W(\Phi , Z) &=& W_{\rm vis}(\Phi) + W_{\rm hidd}(Z).
\label{eq:total-super}
\end{eqnarray}
Here, $\Phi$ and $Z$ denote fields in the visible and hidden sectors,
respectively. We have employed the superspace notation given in \cite{wess-bagger}. 
In the Einstein frame, we see that the \kahler  potential $K$ has the
form of no-scale type\footnote{
The no-scale supergravity \cite{no-scale} adopts a specific form 
$f_{\rm hidd} = Z + Z^{\dagger}$.}
as
\begin{eqnarray}
  K(\Phi , \Phi ^{\dagger}, Z, Z^{\dagger}) = -3\log
    \left(1 - \frac{1}{3}f_{\rm vis}(\Phi , \Phi ^{\dagger}) 
    - \frac{1}{3}f_{\rm hidd}(Z, Z^{\dagger})\right).
\end{eqnarray}
It is a crucial observation in ref.~\cite{kawasaki} that all soft
SUSY-breaking masses for sfermions and $A$ terms in the visible 
sector vanish in the limit of the zero cosmological constant.

In the present model the SU(5)$_{\rm GUT}$ gauge multiplet lives in the
6 dimensional bulk and can couple directly to the hidden-sector field
$Z$ which gives rise to a SUSY-breaking gaugino mass
$m_{\widetilde{G}_5}$. Thus, the bulk SU(5)$_{\rm GUT}$ gaugino 
$\widetilde{G}_5$ may transmit the SUSY breaking on the
hidden 3-brane to the SUSY standard-model sector on our visible 
3-brane ({\it i.e.} gaugino mediation) \cite{kaplan,kawasaki}. 

Let us denote the field responsible for the SUSY breaking $Z_0$ resides 
on a hidden 3-brane separated from our visible 3-brane by extra 
dimensions. The  SU(5)$_{\rm GUT}$ gaugino acquires the SUSY-breaking 
mass $m_{\widetilde{G}_5}$ through the following interaction with $Z_0$; 
\begin{equation}
{\cal S} = \int\int\int d^2\theta \sqrt{-g^{(6)}}\eta
\frac {Z_0(x,y)}{M_*}{\cal W}_\alpha {\cal W}^\alpha 
\delta (y_1-\frac{1}{2}L)\delta (y_2-\frac{1}{2}L)d^4xd^2y,
\end{equation}
where we have taken the position of the hidden brane to be 
$(y_1=L/2,y_2=L/2)$ (an orbifold fixed point). 
Our visible brane is located at the origin of the extra space
$(y_1=0,y_2=0)$ (see Fig.\ref{Fig:T2-Z4-vis-U3}).
The coupling $\eta$ in the above equation is a
dimensionless constant. From this interaction we
obtain the gaugino mass as
\begin{equation}
\frac {m_{\widetilde{G}_5}}{g_{\rm GUT}^2}\simeq \eta 
\frac{\langle F_{Z_0}\rangle}{M_*}, 
\end{equation}
where $\langle F_{Z_0}\rangle$ is the SUSY-breaking $F$-term of the field
$Z_0$. The gravitino mass $m_{3/2}$ is determined as
\begin{equation}
m_{3/2} \simeq  \frac{\langle F_{Z_0}\rangle}{M_{\rm Pl}} 
\end{equation}
and hence we get
\begin{equation}
m_{\widetilde{G}_5}\simeq {g_{\rm GUT}^2}\eta \frac{M_{\rm Pl}}{M_*}m_{3/2}. 
\end{equation}
Since $M_{\rm Pl}/M_* \simeq {\cal O}(10)$, we have relatively large
gaugino mass compared to the gravitino mass if $\eta \simeq {\cal
O}(1)$. We take $\eta \simeq 0.1$ to give the gaugino mass comparable 
to the gravitino mass. This is because the SUSY-invariant 
mass $\mu$ for the Higgs chiral multiplets ($W=\mu H{\bar H}$)
becomes of ${\cal O}(m_{3/2})$ as seen below and phenomenological
analyses in refs.~\cite{kawasaki,kaplan} suggest $m_{\widetilde{G}_5}
\simeq \mu$.

On the other hand, the $\U(3)_{\rm H}$ gauginos remain almost massless,
since the $\U(3)_{\rm H}$ 3-brane is separated from the SUSY-breaking
hidden 3-brane by extra dimensions, and the $\U(3)_{\rm H}$ vector
multiplets do not have a direct coupling to the hidden-sector field $Z_0$.
We neglect their masses in the
present analysis.

Remarkable is that the original SU(5)$_{\rm GUT}\times$U(3)$_{\rm H}$
model predicts the gaugino masses as
\begin{equation}
\frac{m_{\widetilde{G}_3}}{\alpha_3}:
\frac{m_{\widetilde{G}_2}}{\alpha_2}:
\frac{m_{\widetilde{G}_1}}{\alpha_1}
\simeq \left(\frac{m_{\widetilde{G}_5}}{g^2_{\rm GUT}}+
\frac{m_{\widetilde{H}_3}}{g^2_{\rm 3H}}\right):
\frac{m_{\widetilde{G}_5}}{g^2_{\rm GUT}}:
\left(\frac{m_{\widetilde{G}_5}}{g^2_{\rm GUT}}+
\frac{m_{\widetilde{H}_1}}{15g^2_{\rm 1H}}\right), 
\end{equation}
where $m_{\widetilde{H}_3}$ and $m_{\widetilde{H}_1}$ are SUSY-breaking
masses of the SU(3)$_{\rm H}$ and the U(1)$_{\rm H}$ gauginos,
respectively. Thus, we expect, in general, a large
deviation from the GUT mass relation for the SUSY standard-model
gauginos \cite{moroi}. However, we find here $m_{\widetilde{H}_3}\simeq 
m_{\widetilde{H}_1}\simeq 0$, 
which leads to the standard GUT relation,
\begin{equation}
m_{\widetilde{G}_1}\simeq m_{\widetilde{G}_2}\simeq m_{\widetilde{G}_3},
\end{equation}
at the unification scale $M_{\rm GUT}$, where $\alpha_3 \simeq \alpha_2
\simeq \alpha_1$.

With this gaugino masses the SUSY-breaking masses for squarks and
sleptons are generated from one-loop diagrams of intermediate gauginos
as shown in \cite{kawasaki,kaplan}. Since the gauge interactions are
flavor blind, we have degenerate soft SUSY-breaking masses for each 
sfermions with
the same quantum numbers of the standard-model gauge group, and hence
we may suppress the unwanted FCNC's.

We are now at the point to discuss the SUSY-invariant mass of the Higgs
chiral multiplets (called $\mu$ term). It is also important observation
in ref.~\cite{kawasaki} that the $\mu$ term naturally arises from the
\kahler  potential eq.(\ref{kaler}) with
\begin{equation}
 f_{\rm vis}(\Phi, \Phi^{\dagger}) = HH^{\dagger} +{\bar H}{\bar H}^{\dagger}
                               + \tau H{\bar H} + h.c.,
\end{equation}
where $\tau$ is a dimensionless constant of ${\cal O}(1)$.
Notice that this \kahler  potential is invariant of 
the continuous $\U(1)_{\rm R}$ symmetry. After the SUSY breaking,
the superpotential $W$ eq.(\ref{eq:total-super}) should condense 
so that the cosmological constant vanishes. Therefore, the ${\bf Z}_{4{\rm
R}}$ is broken down to the ${\bf Z}_2$ R-parity. In this circumstance,   
the above \kahler  potential induces the $\mu$ term and a SUSY-breaking $B$
term for the Higgs multiplets as 
\begin{equation}
\mu \simeq \tau m_{3/2},~~~B\simeq \mu m_{3/2}.
\end{equation}
Here, $B$ is defined as
\begin{equation}
{\cal L} = B h\bar{h},
\end{equation}
where $h$ and $\bar{h}$ are scalar components of the
Higgs chiral multiplets $H$ and ${\bar H}$, respectively.
This is basically equivalent to the Giudice-Masiero mechanism
\cite{masiero}.

The detailed phenomenology is given in \cite{kawasaki,kaplan,sk-my}. 
We only stress here that the gaugino-mediation model predicts 
the lightest SUSY particle to be a bino or a right-handed slepton 
(a super-partner of right-handed charged lepton) at ${\cal O}(100)$ GeV.  

There may be contact interactions between fields on our
visible 3-brane and on the SUSY-breaking hidden 3-brane, which induce, in
principle, the dangerous FCNC's. Such contact terms
have the following form;
\begin{equation}
{\cal L}_{\rm cont} \sim \frac{e^{- M_*D}}{M^2_*}
\int d^4\theta \Phi \Phi ^{\dagger}Z_0 Z_0^{\dagger},
\end{equation}
which give rise to soft SUSY-breaking masses for squarks and sleptons
that may be generally flavour dependent. 
The distance $D$ between our visible and the hidden 3-brane is
$D= L/\sqrt{2}$ with $M_* L \simeq 20$.
These must be compared with the operators that are induced by gaugino
loops. We easily see that the exponential suppression factor 
$e^{-\frac{1}{\sqrt{2}}M_*L}$ is ${\cal O}(10^{-6})$ in our model, which is 
small enough to satisfy all
constraints from the experimental upper bounds on FCNC's.

\section{String-theory frameworks}
\label{sec:string}

We have described, in section \ref{sec:embedding}, the semi-simple 
unification model based on the $\SU(5)_{\rm GUT} \times \U(3)_{\rm H}$ 
gauge group in terms of a 6-dimensional effective field theory. 
To explain both the quasi-${\cal N}=2$ SUSY in the $\U(3)_{\rm H}$
sector and the disparity of the gauge coupling constants between the
$\SU(5)_{\rm GUT}$ and the $\U(3)_{\rm H}$, we have introduced 
two types of branes.
From the 6-dimensional point of view, one is the $\U(3)_{\rm H}$ 3-brane 
and the other is the $\SU(5)_{\rm GUT}$ 5-brane.
In this section, we embed furthermore our 6-dimensional model into 
string theories, since they may provide natural frameworks
for the brane world. We see that string theories may indeed
accommodate the semi-simple unification model.
In particular, it is extremely interesting that the ${\cal N}=4$ SUSY in 
the 6-dimensional space-time is an automatic consequence of D-brane
structure in the 10-dimensional string theories.
We also show that the string-theory frameworks provide geometric descriptions
for the orbifold group, the discrete low-energy ${\bf Z}_{4}$ R-symmetry 
and the FI term. 
  
For this purpose we first extend our
6-dimensional model to a 10-dimensional theory, where we assume the 
four extra 6789 space dimensions to be compactified smaller than 
the unification scale $M_{\rm GUT}^{-1}$. 
The extension is not unique, since there are many 
10-dimensional string-theoretical configurations which give 
an equivalent 6-dimensional theory after the compactification.
In fact, the string duality \cite{string-duality} connects one 
description to another and those are equivalent 
in the string theory. Therefore, the requirement of reproducing 
the 6-dimensional configuration given in section \ref{sec:embedding} alone 
cannot determine how many dimensions and in what directions the branes
wrap in the compactified four dimensions. In other words, we can choose 
convenient ones depending on the purpose, and we do change descriptions 
place by place in what follows.

The $\U(3)_{\rm H}$ 3-brane that is located in the 6-dimensional bulk is 
naturally lifted up to D3-branes in the type IIB string theory
\cite{notesonDbranes}.
In addition to the D3-branes, we need another type of branes
to hold the $\SU(5)_{\rm GUT}$ on them.
As we mentioned in section \ref{sec:embedding} and in the appendix B, 
the existence of a single type of D-branes breaks half of the SUSY charges.
One might consider that the two different types of D-branes introduced 
above leave 1/4 of the SUSY, and 8 SUSY charges (${\cal N}=2$ in 
4 dimensions) are left unbroken among the 32 SUSY charges of the 
type IIB string theory, but this is not correct in general.
We use D7-branes along with the D3-branes, because it is well known that 
D$p$-D$(p+4)$ brane system keeps the ${\cal N}=2$ SUSY \cite{gk}. We
also explicitly explain this later.
Five D7-branes stretching in 1234567 space directions with three 
D3-branes in 123 space directions lead to a $\U(5) \times \U(3)$ gauge
theory \cite{boundstates}.  
We identify the extra $\U(1)$ on the D7-branes with $B-L$ 
symmetry, and hence it naturally accommodates massive right-handed 
neutrinos.\footnote{
The $\U(5)_{\rm GUT} \times \U(3)_{\rm H}$ is broken down to
$\SU(3)_{\rm C} \times \SU(2)_{\rm L} \times \U(1)_{\rm Y} \times 
\U(1)_{\rm B-L}$ by the condensation in eq.(\ref{vacuum}).}

The field contents of the string-theory D3-D7 brane system are as follows:
D7-D7 open strings provide a $\U(5)_{\rm GUT}$ vector multiplet (${\cal
W}_{\alpha}^{\U(5)_{\rm GUT}},\Sigma,\Sigma',\Sigma''$) of the
6-dimensional (1,1) SUSY (${\cal Q}^{(6)}_{\bf 4_+}=({\cal
Q}^{(4)3,4})$,${\cal Q}^{(6)}_{\bf 4_-}=({\cal Q}^{(4)1,2})$). Bosonic
part of the ${\cal W}^{\U(5)_{\rm GUT}}_\alpha$ and $\Sigma,\Sigma'$ 
arise from 
the fluctuations of strings into 01234567 directions and the scalar component 
of $\Sigma''$ from the fluctuations into 89
directions.\footnote{Here, $\Sigma,\Sigma'$ and $\Sigma''$ represent
adjoint multiplets of $\U(5)_{\rm GUT}$, which consist of ${\bf 24}+{\bf 
1}$ of $\SU(5)_{\rm GUT}$. }
D3-D3 open strings give  a $\U(3)_{\rm H}$ vector multiplet 
(${\cal W}_{\alpha}^{\U(3)_{\rm H}},X''\equiv(X_{\alpha}^{\beta},X_0)$) and
$\U(3)_{\rm H}$ adjoint hypermultiplets $(X,X')$ of
4-dimensional ${\cal N}=2$ SUSY (${\cal Q}^{(4)1},{\cal
Q}^{(4)4}$). Gauge boson of the ${\cal W}^{\U(3)_{\rm H}}_\alpha$ is from the 
fluctuations in 0123 directions in the D3-branes, scalars of the 
$X$ and $X'$ from 4567 directions (transverse to the D3- but
tangential to the D7-branes) and the scalar of the $X''$ from the 89 
directions (transverse to both of the D3- and the D7-branes).
Finally, D3-D7/D7-D3 open strings provide a $\U(3)_{\rm H} \times 
\U(5)_{\rm GUT}$ bi-fundamental hypermultiplet ($Q^k_\alpha,
\bar{Q}_k^\alpha$)  ($\alpha=1,2,3;k=1,\ldots,5$) of the ${\cal N}=2$ 
SUSY on the D3-branes.

All these fields are explicitly described as states of open strings,
 and therefore, the quantum numbers of these fields under 
the space-time symmetry transformations are unambiguously determined.
For later reference, we present their charges under the rotations on the
$x_4$-$x_5$, $x_6$-$x_7$ and $x_8$-$x_9$ planes, respectively,
in Table \ref{u1charges.tbl}.

The description of the brane-world model in the above D3-D7 brane 
system gives an intuitive explanation of
the interaction in eq.(\ref{superpotential}),
\begin{equation}
 W = \lambda \bar{Q}_k X'' Q^k.
\end{equation}
Namely, a displacement of the D3-branes away from the D7-branes ({\it i.e.}
non-zero expectation value of the scalar of $X''$) leads to massive open
strings ($Q^k,\bar{Q}_k$) stretching between the D3- and D7-branes.

Next, let us discuss SUSY and R-symmetries.
We take the T-duality of the D3-D7 brane system 
along $x_8$ and $x_9$ directions and we discuss in terms of D5-D9 
brane system for a while.

We first focus on the $\U(5)_{\rm GUT}$ gauge theory.
The D9-brane breaks half of the 32 SUSY charges of the type IIB string theory,
and hence ${\cal N}=4$ SUSY is left unbroken in the sense of 4-dimensional
effective theory. 
The condition for unbroken SUSY charges is given as follows.  
The SUSY charges of the type IIB string theory are represented as
two Majorana-Weyl spinors ${\cal Q}^{(10)1}$ and ${\cal Q}^{(10)2}$ 
with a common chirality (both ${\bf 16_+}$ of the $\SO(9,1)$).
Each of them has $16$ real components.
We combine them into a complex Weyl spinor
${\cal Q}^{(10)}\equiv {\cal Q}^{(10)1}+i{\cal Q}^{(10)2}$.
The condition is written as
\begin{equation}
{\cal Q}^{(10)}=\Gamma_{d=10}^{0123456789}{\cal Q}^{(10)c},
\label{realitycond}
\end{equation}
where the ${\cal Q}^{(10)c}$ in the right-hand side is the charge
conjugation of the ${\cal Q}^{(10)}$,
$\Gamma^\mu_{d=10}$ gamma matrices of the $\SO(9,1)$, and
$\Gamma^{01\cdots 9} \equiv \Gamma^0\Gamma^1 \cdots \Gamma^9$. (See
appendix A for other notations.)

It is known that the type IIB string theory compactified into the 
6 dimensions has maximally $\SL(5,{\bf R})$ symmetry \cite{HullTownsend}.
This symmetry is generated by rotations in the compactified four
dimensions and other internal transformations.\footnote{
Precisely speaking, this symmetry is broken down to a discrete subgroup 
$\SL(5,{\bf Z})$ due to finite-size effects of the compactification.
The spectrum of the Kaluza-Klein modes on the ${\bf
T}^4$ torus
depends on the shape of the torus and it does not respect the rotational 
symmetry. This fact plays an important role in our model and we will 
return to this point. However, we discuss here a symmetry for massless 
modes and neglect the symmetry breaking.}
If we introduce $\U(5)_{\rm GUT}$ D9-branes,
the above internal symmetry is broken by the brane-bulk interaction term
in the D-brane action and the $\SL(5,{\bf R})$ symmetry 
reduces to purely a rotational symmetry of the internal space, that is, to 
the $\SO(4)_{6789}$. 
The SUSY charges transform as spinors of the $\SO(4)_{6789}$, since they 
form a spinor of the $\SO(9,1)$. Therefore, the $\SO(4)_{6789}$ is regarded
as an R-symmetry, just as the $\SO(2)_{45}$ is regarded as an R-symmetry 
after the dimensional reduction to 4 dimensions as shown in section
\ref{sec:embedding}. 
The decomposition of SUSY charge ${\bf 16_+}$ of the $\SO(9,1)$ into
the representations of a subgroup
$\SO(5,1) \times \SO(4)_{6789}$ is
\begin{equation}
{\bf 16_+}\rightarrow
({\bf 4_+},({\bf 2},{\bf 1}))+({\bf 4_-},({\bf 1},{\bf 2})).
\label{421412}
\end{equation}
Each term in the right-hand side corresponds to the 
SUSY charges $Q_{{\bf 4_+},A}^{(6)}$ and $Q_{{\bf 4}_-,B}^{(6)}$, 
respectively.
The rotational symmetry $\SO(4)_{6789}$ of the internal space 
is regarded as the $\SU(2)_{\bf 4_+} \times \SU(2)_{\bf 4_-}$ R-symmetry 
of the (1,1) SUSY in the 6-dimension discussed in section \ref{sec:embedding}.
Due to the reality condition eq.(\ref{realitycond}), these SUSY charges
are pseudo-Majorana-Weyl (see appendix A).

The $\U(5)_{\rm GUT}$ gauge theory on the D9-branes has ${\cal N}=4$
SUSY in 4-dimensional effective theory as stated before. In this description, 
the rotational symmetry $\SO(6)_{456789}$ of the internal 6-dimensional 
space is identified with the R-symmetry $\SU(4)_{\rm R}$ of the ${\cal N}=4$
SUSY theory discussed in section \ref{sec:embedding}.
First, we show that the action of the $\SO(6)_{456789}$ on the ${\cal
N}=4$ SUSY charges, gauginos, and scalars are the same as those of the
$\SU(4)_{\rm R}$ given in section \ref{sec:embedding}.
(${\bf 16_+}$)-spinor (to which the SUSY-charge ${\cal Q}^{(10)}$ belongs)
and (${\bf 16_-}$)-spinor (to which the $\U(5)_{\rm GUT}$ adjoint
gauge fermions and the SUSY-transformation parameters belong) are 
decomposed as
\begin{eqnarray}
{\bf 16_+} &\rightarrow &({\bf 2}_L, {\bf 4}) +({\bf 2}_R,{\bf 4}^*),
\label{eq:decomp} \\
{\bf 16_- } &\rightarrow &({\bf 2}_L, {\bf 4}^*) +({\bf 2}_R,{\bf 4}), 
\end{eqnarray}
of the $\SO(3,1) \times \SO(6)_{456789}$.
Two terms in the right-hand side of eq.(\ref{eq:decomp}) correspond to 
${\cal Q}_\alpha^{(4)a}$ and $\bar{\cal Q}^{(4)\dot\alpha}_a$, 
respectively.\footnote{The reality condition eq.(\ref{realitycond}) 
implies that ${\cal Q}^{(4)a}_\alpha$'s and $\bar{{\cal
Q}}^{(4)}_{\dot{\alpha},a}$'s  are complex conjugate of each other.}
Thus, we have ${\cal N}=4$ SUSY charges belonging to ${\bf 4}$-spinor of
the $\SO(6)_{456789}$ ({\it i.e.} ${\bf 4}$ of the $\SU(4)_{\rm R}$).
It is easy to see here that the $\U(5)_{\rm GUT}$ adjoint
gauge fermions $\chi_a$ belong to ${\bf 4}^*$-spinor of the
$\SO(6)_{456789}$, and the $\U(5)_{\rm GUT}$ adjoint scalars 
$\sigma,\sigma',\sigma''$ to vector of the $\SO(6)_{456789}$ (by definition) 
as in eq.(\ref{eq:so(6)-vector}). Therefore, the $\SO(6)_{456789}$ is
really the $\SU(4)_{\rm R}$ symmetry considered in section
\ref{sec:embedding}. 
Furthermore, we explicitly find, from the Table \ref{u1charges.tbl} which 
shows the charges of fields for the maximal torus $\SO(2)_{45}\times
\SO(2)_{67}\times \SO(2)_{89}$ of this $\SO(6)_{456789}$, 
that the charges of the 2nd-lowest component fields are smaller than those of 
the lowest components by the charge of the SUSY generator ${\cal
Q}^{(4)1}$. Therefore, each of these $\SO(2)$ symmetries can be
identified with the usual R-symmetry in ${\cal N}=1$ SUSY theories.  

Now, let us consider the SUSY breaking
${\cal N}=4\rightarrow{\cal N}=2$.
The SUSY breaking due to the presence of the $\U(3)_{\rm H}$ D5-branes 
along $012389$ directions is represented by the following constraint 
on the SUSY charges:
\begin{equation}
{\cal Q}^{(10)}=-\Gamma^{012389}_{d=10}{\cal Q}^{(10)c}.
\label{d5susy}
\end{equation}
Combining eq.(\ref{d5susy}) and eq.(\ref{realitycond}),
we obtain a holomorphic constraint on the ${\cal Q}^{(10)}$:
\begin{equation}
{\cal Q}^{(10)}=-\Gamma^{4567}_{d=10}{\cal Q}^{(10)}.
\label{4567susy}
\end{equation}
With the convention of gamma matrices described in the appendix A,
where generators of the Cartan of the $\SO(6)_{456789}\simeq \SU(4)_{\rm 
R}$
are given by\footnote{
The $T^{mn}_{(6)}$ is the generator of rotation on $x_m$-$x_n$ plane.}
\begin{eqnarray}
T^{45}_{(6)}|_{\bf 4}\equiv \frac{i}{2} \gamma_{(6)}^{45}|_{\bf 4} &=&
\frac{1}{2} \diag(-1,-1,+1,+1),\nonumber\\
T^{67}_{(6)}|_{\bf 4}\equiv \frac{i}{2} \gamma_{(6)}^{67}|_{\bf 4} &=&
\frac{1}{2} \diag(-1,+1,-1,+1),\nonumber\\
T^{89}_{(6)}|_{\bf 4}\equiv \frac{i}{2} \gamma_{(6)}^{89}|_{\bf 4} &=&
\frac{1}{2} \diag(-1,+1,+1,-1),
\label{3gen}
\end{eqnarray}
the condition eq.(\ref{4567susy}) is rewritten as
\begin{equation}
\left(\begin{array}{cccc}
1 \\
&-1 \\
&&-1 \\
&&&1
\end{array}\right)
\left(\begin{array}{c}
{\cal Q}^{(4)1} \\ {\cal Q}^{(4)2} \\ {\cal Q}^{(4)3} \\ {\cal Q}^{(4)4}
\end{array}\right)
=\left(\begin{array}{c}
{\cal Q}^{(4)1} \\ {\cal Q}^{(4)2} \\ {\cal Q}^{(4)3} \\ {\cal Q}^{(4)4}
\end{array}\right).
\end{equation}
From this equation, we see that ${\cal Q}^{(4)2}$ and ${\cal Q}^{(4)3}$ 
are broken and the ${\cal N}=2$ SUSY generated by 
${\cal Q}^{(4)1}$ and ${\cal Q}^{(4)4}$ is realized on the D5-branes, as 
stated before. 

Because the D5-branes are wrapped only along two of six
internal directions ({\it i.e.} $x_8$ and $x_9$ directions),
the $\SO(6)_{456789}$ symmetry is actually broken into
$\SO(4)_{4567} \times \SO(2)_{89}$, which is the same as the $\SU(2)_{\rm R} 
\times \SU(2)_{\rm F} \times \U(1)_{\rm R}$ in 
eq.(\ref{eq:embed-U(1)89*SU(2)R*SU(2)L}).
What is important here is that the R-symmetries relevant to the
phenomenology ({\it i.e. }$\SU(2)_{\rm R}$ and $\U(1)_{\rm R}$) have 
geometrical interpretations. The $\U(1)_{\rm R}$ symmetry is   
the space-rotational symmetry $\SO(2)_{89}$ in the compactified internal 
$x_8$-$x_9$ plane. The $\SU(2)_{\rm R}$ and $\SU(2)_{\rm F}$ are
$\SU(2)$ rotation symmetries of complex coordinates
$(x_4+ix_5,x_6-ix_7)$ and $(x_4+ix_5,x_6+ix_7)$, respectively. This can
be understood by observing how the $\SU(2)_{\rm R}$ and $\SU(2)_{\rm F}$ act
on the $\sigma,\sigma',\sigma^*$ and $\sigma^{'*}$ (see  
eq.(\ref{eq:N4-scalar-multiplet}) and
eq.(\ref{eq:embed-U(1)89*SU(2)R*SU(2)L})).

Finally, we reinterpret the ${\bf Z}_4$ orbifolding
of ${\bf T}^2$ in the 6-dimension as one of ${\bf T}^6$.
In the $6$ dimensions, the ${\bf Z}_4$ generator in eq.(\ref{Z4-generator})
is a $- \pi/2$ rotation of the $x_4$-$x_5$
plane accompanied by an R-rotation.
Because an R-rotation is a rotation on the internal space 
in the string framework and the $\SU(4)_{\rm R} \simeq \SO(6)_{456789}$ 
is nothing but the rotation in $456789$ directions,
we can consider the orbifolding in a totally geometrical way.
We find that the eq.(\ref{Z4-generator}) is identified with the following 
rotation:
\begin{equation}
(x_4 + i x_5) \rightarrow e^{-\pi i/2} (x_4 + i x_5),\quad
(x_6 + i x_7) \rightarrow e^{-\pi i/2} (x_6 + i x_7),\quad
(x_8 + i x_9) \rightarrow e^{\pi i} (x_8 + i x_9).
\label{z4rot}
\end{equation}
Indeed, with the convention of the $\gamma$-matrices in eq.(\ref{3gen}),
the ${\bf 4}$-spinor representation matrix of the rotation eq.(\ref{z4rot})
is given by
\begin{equation}
\exp\left(\frac{-\pi i}{2}T^{45}_{(6)}\right)
\exp\left(\frac{-\pi i}{2}T^{67}_{(6)}\right)
\exp\left(\pi i T^{89}_{(6)}\right)|_{\bf 4}=
\diag(1,i,i,-1),
\end{equation}
and this is identical with the transformation eq.(\ref{Z4-generator}).
Therefore, the invariance under this orbifold transformation allows
only one SUSY charge ${\cal Q}^{(4)1}$ to be left unbroken as we have shown 
in section \ref{sec:embedding}.

We obtain a fully geometrical picture of the orbifolding in
the 10-dimensional description. The geometry of the
compactified manifold in the extra 6 dimensions must be consistent with the
orbifold-group symmetry.
Two ${\bf T}^2$'s along $x_4$-$x_5$ and $x_6$-$x_7$ directions should 
be square because the orbifold-group action rotates both $x_4$-$x_5$
and $x_6$-$x_7$ by $-\pi/2$.
On the contrary, the torus along $x_8$-$x_9$ directions can be of
generic shape, because any toroidal compactification preserve  
${\bf Z}_2$ symmetry generated by the angle-$\pi$ rotation.
Taking account that the $\SO(2)_{89}$ also has spinor representations,
we see that the unbroken symmetry of the torus in $x_8$-$x_9$ direction 
is really a ${\bf Z}_4$.
This is identified with the ${\bf Z}_{4{\rm R}}$ symmetry.\footnote{
This ${\bf Z}_4$ R-symmetry is a gauge symmetry originating from the
$\SO(9,1)$.}
Comparing Table \ref{u1charges.tbl} with Table \ref{tab:R-charge},
we find $Q^k$, $\bar{Q}_k$ and $X''(\equiv(X^\alpha_\beta , X_0))$ have 
desirable charges under this symmetry.
Notice that the ${\bf Z}_{4}$ R-symmetry is a natural
consequence of a theory given by a toroidal compactification in 
a higher dimensional theory, and the 4-dimensional effective theory 
discussed in section \ref{sec:5-3} really has this ${\bf Z}_4$ R-symmetry. 
It is extremely encouraging that the discreteness of the R-symmetry 
(and more precisely, the reason why the R-symmetry is broken down 
to ${\bf Z}_4$) is explained naturally when we extend the 6-dimensional 
effective theory into this 10-dimensional theory with toroidal 
compactification.

 
There is another important feature of our model;
that is, the FI F-term parameter $V^2$ in eq.(\ref{superpotential}).   
It is possible to put the FI F-term by hand in the $\U(3)_{\rm H}$-brane
superpotential, if we regard this 10-dimensional theory just as a field
theory.  
In string theories, FI parameters in the ${\cal N}=2$ SUSY 
D$p$-D($p+4$)-brane system may be explained by a string $B$-field 
background \cite{B-field-FI} as we see below.

The 4-dimensional ${\cal N}=2$ SUSY-transformation law of gauginos is given
by
\begin{equation}
\delta \psi \sim (F_{ij}\gamma^{ij}_{d=4}+\xi_a \tau_a)\epsilon
+(\mbox{terms containing scalar fields}),
\label{eq:deltapsi-general}
\end{equation}
where the gauginos $\psi$ and the transformation parameter $\epsilon$ are
$\SU(2)_{\rm R}$ doublets, $\tau_a$ the Pauli matrices which act
on $\SU(2)_{\rm R}$ doublets, and $\xi_a$ ($a=1,2,3$) are $\SU(2)_{\rm
R}$ triplet FI parameters that break the $\SU(2)_{\rm R}$ symmetry.
$\xi_3$ is FI D-term parameter, and $\xi_1 + i\xi_2$ the FI F-term
parameter that corresponds to the $V^2$ in the superpotential 
eq.(\ref{superpotential}). ($V$ is of order of $M_{\rm GUT} \simeq 
\tilde{L}^{-1}$.) On the other hand, the ${\cal N}=2$ SUSY
transformation of gauginos on D-branes is given by 
\begin{equation}
\delta \psi^A = \frac{-1}{2} {\cal F}_{ij}\gamma^{ij}_{d=4}\epsilon_B 
                                                         \epsilon^{AB} 
                +\frac{-1}{2} {\cal F}_{mn}\Gamma^{mn}\epsilon_B \epsilon^{AB}
                - i \epsilon^{AB} \left( \begin{array}{cc}
		  D & -\sqrt{2}i F^* \\ \sqrt{2}i F & -D 
			\end{array}\right)_{B}^{~~~C}\epsilon_C 
                 + \cdots,
\label{eq:deltapsi-string}
\end{equation}
where $\epsilon^{AB}=\left(\begin{array}{cc}
		      &1 \\-1 &
			   \end{array}\right)$, ${\cal F}_{\mu\nu}=
F_{\mu\nu}+B_{\mu\nu}$ ($B_{\mu\nu}$ the string $B$-field) and
$(-\sqrt{2}\Im F,\sqrt{2}\Re F,\\ D)$ are the $\SU(2)_{\rm R}$ triplet 
auxiliary fields in the ${\cal N}=2$ vector multiplet.
Notice that, in the {\it full} string theory on D-branes, a gauge 
field $F_{\mu\nu}$ and the $B$-field always appear in a combination
$F_{\mu\nu}+B_{\mu\nu}$. If the $B$-field has background values, the 2nd 
term in eq.(\ref{eq:deltapsi-string}) gives rise to a non-linear shift 
SUSY transformation, which is nothing but the 2nd term in
eq.(\ref{eq:deltapsi-general}) ({\it i.e. }the FI contributions).
Therefore, it is possible that the FI-term is induced by the $B$-field
background in the extra space dimensions.    
The $B$-field has six components in four perpendicular directions to the 
D5-branes. They are decomposed into
anti-self-dual part ($\SU(2)_{\rm F}$ triplet)
$(B_{56}-B_{47},B_{64}-B_{67},B_{45}-B_{67})$
and self-dual part ($\SU(2)_{\rm R}$ triplet)
$(B_{56}+B_{47},B_{64}+B_{57},B_{45}+B_{67})$.
The latter triplet is identified with the FI parameters $\xi_a$.
To determine the correspondence between three FI-parameters and three 
self-dual components of the $B$-field, we examine their behavior
under the $\U(1)$ R-symmetry generated by $T^{45}_{(6)}+T^{67}_{(6)}$, 
which is the diagonal $\U(1)$ subgroup of the $\SU(2)_{\rm R}$. 
Among the $\SU(2)_{\rm R}$ triplet
$(B_{56}+B_{47},B_{64}+B_{57},B_{45}+B_{67})$, 
only one component $B_{45}+B_{67}$ is invariant under this $\U(1)$.
Thus, we can conclude that $B_{45}+B_{67}$ is identified with the D-term FI 
parameter $\xi_3$ and other two components correspond to the 
F-term FI parameter $\xi_1 + i\xi_2$.\footnote{Another derivation of
this explicit correspondence between the $B$-field vev's and the 
FI parameters is given in the appendix C, where ${\cal F}_{mn}\Gamma^{mn}$ 
is explicitly calculated using our gamma-matrix convention.}
If the latter components take appropriate expectation values,
we can obtain nonzero value of parameter $V$
in the superpotential in eq.(\ref{superpotential}).
 
Up to now, we have seen that the SUSY, R-symmetry, and some part 
of the field contents in the 6-dimensional brane world scenario are
naturally reproduced in the string framework.
If we proceed further, however, we come up against some problems.

The first one is about the Ramond-Ramond (RR) charge cancellation 
of the branes. For D5-branes, all the four transverse directions 
are compactified. In such a situation, the total charge of D5-branes 
should vanish because the flux has nowhere to go. This is also the case 
for D9-branes. D9-branes fill up the whole 10-dimensional space-time and
the flux of the D9-brane charge has really nowhere to go.
This severely restricts the number of branes, as we see below.
 
First, let us discuss the cancellation of the RR charges of the D9-branes.
Because each of the D9-branes has the same charge, we cannot put any 
D9-brane unless we put other kind of objects with opposite sign of 
the RR charge. For example, we can use an orientifold 9-plane (O9-plane) 
to cancel the charge of D9-branes without breaking SUSY
\cite{notesonDbranes,orientifold}.
An orientifold $p$-plane (O$p$-plane) is a fixed point of a specific 
${\bf Z}_2$-orbifolding of a torus that flips $(9-p)$ space directions. 
In the case of an O9-plane, the ${\bf Z}_2$-flip is an internal operation 
reversing the orientation of strings, and this ``fixed O9-plane'' 
fills up the whole space-time as the D9-branes do. The charge of an 
O9-plane is $-32$ \cite{notesonDbranes,orientifold}, if we normalize 
the RR charge of a D9-brane to be unity. Therefore, if we use an O9-plane 
to cancel the D9-brane charges, we should put 32 D9-branes in a torus.
The type IIB string theory with an O9-plane and $32$ D9-branes
(compensating the RR charge of the O9-plane) is nothing but the type I 
string theory. This theory has an $\SO(32)$ gauge group in the 
10-dimensional bulk.

This $\SO(32)$ gauge group is much larger than what we want 
({\it i.e.} $\U(5)_{\rm GUT}$), and hence this symmetry must be somehow 
broken down. We always have adjoint Higgs fields obtained from the 
$\SO(32)$ gauge field by the compactification. If these Higgs fields
acquire appropriate vev's, the gauge group would be broken down to a smaller 
group. However, this mechanism does not work straightforward in our model.
This is because the massless modes of the adjoint Higgs fields
are eliminated by the orbifolding.\footnote{ 
The orbifolding here is ${\bf Z}_4 \times$``${\bf Z}_2$''$\times 
$``${\bf Z}_2$'', where the last two (``${\bf Z}_2$'')'s are the space
reflections associated with the O$p$-plane and O$(p+4)$-plane in the
D$p$-D$(p+4)$ brane system. In the D5-D9 description, the last 
``${\bf Z}_2$'' is associated with the O9-plane and does not correspond
to an action on the real space. The ``orbifold group'' is now larger, 
and it seems that the unbroken ${\cal N}=1$ SUSY discussed in 
section \ref{sec:embedding} is lost. However,
the unbroken SUSY condition from the ``${\bf Z}_2$'' orbifolding
associated with the O$p$-planes are always equivalent to the condition
form D$p$-branes, and hence there occurs no further breaking of the SUSY.
 We briefly explain this in the appendix D.}
Thus, space-independent vev's of these Higgs fields are forbidden.
Then, how about the condensation of the Kaluza-Klein modes?
There are many Kaluza-Klein modes which are not projected out 
by the orbifolding. Although condensation of massive Kaluza-Klein 
modes is impossible within small perturbations from the
$\SO(32)$-symmetric vacuum, scalar fields which are massive at 
symmetric vacuum generically have other degenerate vacua 
(where the gauge symmetry is broken) distant from the symmetric vacuum 
in ${\cal N}=1$ SUSY field theories.\footnote{
For example, an ${\cal N}=1$ SUSY theory described by a superpotential 
$W= m\Phi^2 +\lambda \Phi^3 $ satisfies the property discussed in the
text. Notice that the SUSY governing interactions of the $\U(5)_{\rm GUT}$
multiplets is not ${\cal N}=4$ but ${\cal N}=1$.}  
Therefore, we cannot reject the possibility that some unknown 
non-perturbative effects deform the potential of the Kaluza-Klein 
modes and as a result, the Higgs fields may obtain space-dependent vev's,
which break the gauge symmetry $\SO(32)$ down to the desired one.

The charge cancellation for D5-branes also requires the existence of
orientifold 5-planes (O5-planes) and a large gauge group.
We should break the gauge symmetry on the D5-branes again.
It is much easier to obtain the desired gauge group than in 
the D9 case.
This is because the Higgs fields on the D5-branes localize on the
D5-branes which can be free from the orbifold fixed points, whereas the
D9-branes always include fixed points on them. 
Thus, we can use the constant vev's of the adjoint Higgs fields to break 
the gauge symmetry on the D5-branes.

However, the RR-charge cancellation by the O5-planes 
does not work in the present model. Since we put in the bulk 
three D5-branes on which the $\U(3)_{\rm H}$ sector resides, there exist 15
images under the ${\bf Z}_4 \times$``${\bf Z}_2$''$\times$``${\bf
Z}_2$'', and hence we have 48 D5-branes. This problem can be, however,
easily solved by replacing the orbifold group ${\bf Z}_4$ in 
eq.(\ref{Z4-generator}) by a ${\bf Z}_2$ whose 
generator is now given by diag$(1,-1,1,-1)$. Owing to the orientifold 
reflection ``${\bf Z}_2$''$\times$``${\bf Z}_2$'' essential features 
remain unchanged.

The condensation of Kaluza-Klein modes
(non-constant modes) seems to cause the breakdown of the SUSY.
However, field-independent contribution to the SUSY transformation from 
the derivatives (Kaluza-Klein momenta) of these condensations
can be canceled by those from derivative terms of other background fields 
(such as the metric, $B$-field, which we discuss later, gauge fields and 
scalars). Indeed, once the Kaluza-Klein condensation takes place, 
the uniformity and flatness are lost in the background space-time of the
compactified manifold. Background metric with nonzero derivative is
rather natural.
Although this picture is far beyond the small perturbation from
the flat-torus-compactification picture,  
once we accept it, we may be able to resolve other 
problems of the D-brane construction of our model.

One relevant problem is
that we have two unwanted $\SU(3)_{\rm H}$ adjoint chiral multiplets
$X$ and $X'$ on the D5-branes.
They are NG-bosons (and their fermionic partners) associated with the
breaking of the parallel transport symmetry due to the
existence of the D5-branes
and these bosons represent fluctuations of the D5-branes.
To eliminate these massless fields,
we have to fix the D5-branes at some point in the compactified manifold.
If the background space-time is not uniform,
this problem may be solved naturally.\footnote{For example, see 
\cite{imamura}.}

Furthermore, the condensation of Kaluza-Klein modes might resolve a
problem concerning the FI term. We go back to the D3-D7 description
again, taking the T-duality.\footnote{The T-dual of the orbifold 
${\bf T}^6/{\bf Z}_4$ and orientifold ${\bf T}^6/\left({\bf Z}_2 \times
\right.$``${\bf Z}_2$''$\times$``${\bf Z}_2$''$\left.\right)$ are 
${\bf T}^6/{\bf Z}_4$ and ${\bf T}^6/\left({\bf Z}_2
\times\right.$``${\bf Z}_2$''$\times$``${\bf Z}_2$''$\left.\right)$, 
where the ``${\bf Z}_2$''$\times$``${\bf Z}_2$'' is now the ${\bf Z}_2$ 
flips associated with the O3- and O7-planes, respectively.} 
Among three components of the $B$-field corresponding to the three 
FI parameters $\xi_a ~(a=1,2,3)$, only the D-term parameter 
$\xi_3 \sim B_{45}+B_{67}$ is invariant under the orbifolding 
eq.(\ref{z4rot}). Therefore, if we assume a constant $B$-field, 
we cannot obtain nonzero FI F-term parameter $(\xi_1+i\xi_2) \sim 
V^2 $ in the superpotential eq.(\ref{superpotential}). However, once the 
uniformity of the background space-time is broken, the presence of a 
space-dependent $B$-field is quite natural. Therefore, there is no 
reason to forbid the emergence of the nonzero $\xi_1+i\xi_2$ parameter on the 
$\U(3)_{\rm H}$ D3-branes and we may expect the desirable value 
$|\xi_1 + i \xi_2| \sim |V|^2 \sim M_{\rm GUT}^2$.

The D-brane construction has another kind of difficulty.
Unlike the gauge fields and matter fields $X''$, $Q_\alpha^k$ and
$\bar{Q}^\alpha_k$, 
it is not straightforward to accommodate the anti-symmetric tensor ${\bf
10}$'s of the $\SU(5)_{\rm GUT}$ in the string-framework brane world.
However, there have been trials in how to embed the standard model or
the anti-symmetric tensor in the string-framework brane world
\cite{sm-A,a-sym-C,5+10-L}. 
It is possible, for example, that the ${\bf 10}$'s arise from five 
D3-branes located at the visible-sector fixed point via an orbifolding
whose action on the space-time is accompanied by a rigid gauge transformation.
There, the $\U(5)_{\rm GUT}$ is the diagonal subgroup of the fixed-point 
D3-$\U(5)$ and the D7-$\U(5)$. We will investigate this possibility in
future publication \cite{future}.  
Similarly, the origin of fields $Q^6_\alpha$ and $\bar{Q}_6^\alpha$
in the $\U(3)_{\rm H}$ sector is not clear.
Because they are neutral with respect to the $\SU(5)_{\rm GUT}$,
we cannot simply identify them to
D3-D7 (or D5-D9 in the T-dual picture) open strings.
One possibility is that they are provided by strings stretching between 
D3-branes and the orbifold fixed point $F_v$.
Because one of two end points are on the D3-branes,
these modes would belong to the fundamental representation of the 
$\SU(3)_{\rm H}$.
To make $Q^6_\alpha$ and $\bar{Q}^\alpha_6$ light,
the D3-branes should be sufficiently close to the fixed point.
This seems to induce a side effect that modes of open strings stretching
between a D3-brane and its mirror image have the same order of masses
as the mass of the  $Q^6_\alpha$ and $\bar{Q}^\alpha_6$.
However, if the background manifold is deformed as we mentioned above,
these modes may acquire masses by interactions with the background field
and it may be possible to obtain only $Q^6_\alpha$ and $\bar{Q}_6^\alpha$.

All arguments above depend on the string dynamics, which seems, however, 
very unclear at the present stage of understanding the string theories,
and it is beyond the scope of this paper. Therefore, we consider that
more intense studies on the string dynamics are highly desired for a
fully (phenomenologically and theoretically) consistent
string-unification model.

\section{Discussion and conclusions}

The SUSY $\SU(5)_{\rm GUT} \times \U(3)_{\rm H}$ model
\cite{yanagida,hy,iy} is quite an attractive candidate for an effective
theory at the unification scale $M_{\rm GUT} \simeq 10^{16}$GeV, since
it naturally explains light Higgs doublets keeping the success of the
conventional SUSY grand unified theories. The phenomenological study 
of the original model suggests a number of mysterious, but interesting 
features. 
In particular, the approximate unification of three gauge coupling
constants $\alpha_3, \alpha_2$ and $\alpha_1$ implies that the
hypercolor $\U(3)_{\rm H}$ gauge interactions are in a strong coupling
regime. This requirement leads us to propose a relatively lower cut-off
scale $M_* \simeq 10^{17}$GeV, since the abelian $\U(1)_{\rm H}$ gauge
coupling constant blows up below the Planck scale $M_{\rm Pl} \simeq 2
\times 10^{18}$GeV. Thus, we are forced 
to embed the original semi-simple $\SU(5)_{\rm GUT} \times 
\U(3)_{\rm H}$ model in the brane world in a higher dimensional
space-time, in which the Planck scale $M_{\rm Pl}$ is merely an 
effective one and the fundamental scale is rather the cut-off
scale $M_*$.

In this paper, we assume a 6-dimensional space-time and consider 
that the two extra 
dimensions are compactified into the orbifold ${\bf T}^2/{\bf Z}_4$
which has two ${\bf Z}_4$ fixed points. The standard quark, lepton 
and Higgs chiral multiplets are assumed to reside on a 3-brane at 
one of the orbifold
fixed points. We put the $\SU(5)_{\rm GUT}$ gauge vector multiplet in
the 6-dimensional bulk. On the other hand, the hypercolor 
$\U(3)_{\rm H}$-sector fields are assumed to reside on a 3-brane in the
6-dimensional bulk. A crucial assumption is the ${\cal N}=4$ SUSY in the 
6-dimensional bulk, which is broken down to ${\cal N}=1$ SUSY by the
${\bf Z}_4$ orbifolding. This ${\cal N}=4$ SUSY is very important to preserve the
quasi-${\cal N}=2$ structure in the $\U(3)_{\rm H}$ sector that is one
of the mysteries in the original $\SU(5)_{\rm GUT} \times \U(3)_{\rm H}$  
unification model.

The present brane-world scenario is very interesting, since it may 
explains naturally the various mysterious features in the original model. 
However, it produces another mystery that is the ${\cal N}=4$ SUSY in
the 6-dimensional bulk. The necessity of ${\cal N}=4$ SUSY leads us again 
to postulate further a higher dimensional space-time, that is a
10-dimensional one. It may be surprising that the phenomenological
consideration suggests strongly some 10-dimensional supergravity theory.
Clearly, the most attractive candidate for it is the superstring
theories. We find, in this paper, that the type IIB string theory with
D3-D7 brane structure may accommodate our 6-dimensional brane-world
model, where the ${\cal N}=4$ SUSY is the automatic prediction of the
theory. The phenomenologically required natures of the 
$\U(3)_{\rm H}$ 3-brane fields are naturally explained in the
type IIB string framework, which seems a miraculous success in this 
framework, since those natures are just imposed by hand in the original
unification model. 
However, there are many unsolved problems in the present string-theory
framework as pointed out in section \ref{sec:string}. In particular, it
is not easy to have the chiral nature ${\bf 5}^* + {\bf 10}$ of the 
quark and lepton multiplets. The RR charge cancellation suggests 
an $\SO(32)$ gauge symmetry on D9-branes (T-dual to the D7-branes) 
instead of the $\U(5)_{\rm GUT}$ and hence we have to make a dynamical 
assumption that the $\SO(32)$ is broken down to $\U(5)_{\rm GUT} \times 
$ (something). However, it does not seem obvious to justify such a 
dynamical breaking at the present knowledge of the string 
theories.\footnote{Another way of solving this problem
will be given by postulating non-compactified $8$-$9$ directions which 
may form a anti-de Sitter space. We will discuss this possibility in 
the future publication \cite{future}.}
Even so, we view the fact that the ${\cal N}=4$ SUSY in the
6-dimensional bulk, the discrete ${\bf Z}_4$ R-symmetry, the FI term and 
an adequate matter content of the $\U(3)_{\rm H}$ sector can be naturally
obtained in the string framework as quite encouraging, and hence
consider that the present string-unification model
deserves further investigations.

Furthermore, it is very interesting that some of the predictions of the
present model are experimentally testable in future. One is
the $\U(1)$ factor associated with D7-branes, which may be regarded as
the $B-L$ symmetry. The presence of the $\U(1)$ implies massive neutrinos 
\cite{see-saw} that are indeed almost confirmed by neutrino oscillation
experiments \cite{oscill}. Another is the GUT-like relation of gaugino
mass as discussed in section \ref{sec:gaugino-med}, which may be tested
in near future collider experiments. As we discussed in section
\ref{sec:gaugino-med} the gaugino-mediation model of the SUSY breaking
is the most natural in the present brane-world model, 
which may be also tested in future experiments, since it
predicts a very peculiar spectrum for many SUSY particles
\cite{kawasaki,kaplan,sk-my} 
\section*{Acknowledgments}
The authors (T.W. and T.Y.) are grateful to K.-I. Izawa for discussion. 
This work was partially supported by ``Priority Area: Supersymmetry
and Unified Theory of Elementary Physics ({\#} 707)'' (T.Y.) and 
the Japan Society for the Promotion of Science (T.W.).
%

\appendix
\section{Conventions of gamma matrices and spinors}

\subsection{The $\SO(3,1)$ gamma matrices and spinors}
The metric convention we take in this paper is $\eta_{\mu\nu}
=\diag (-1,+1,\cdots,+1)$. 

4-dimensional gamma matrices $\gamma_{d=4}^i ~(i=0,1,2,3)$ 
are given by $4 \times 4$ matrices 
\begin{equation}
 \gamma^i_{d=4} \equiv \left(\begin{array}{cc}
			& i \sigma^i \\ i \bar{\sigma}^i &
			     \end{array}\right)  \equiv
\left(
   \left(\begin{array}{cc}
	          & i {\bf 1} \\
        i {\bf 1} & 
	 \end{array}\right),
   \left(\begin{array}{cc}
                  & i \vec{{\bf \tau}}\\
       -i \vec{{\bf \tau}} &
         \end{array}\right)
\right),
\end{equation}
where $\vec{{\bf \tau}}$ are the Pauli matrices.
Using $4 \times 4$ matrices $B_{d=4}$ and $C_{d=4}
$\begin{eqnarray}
B_{d=4} \equiv i \gamma^0\gamma^1\gamma^3, & \quad & 
B_{d=4}^{-1}\left(\gamma^{ij}\right)^* B_{d=4} = \gamma^{ij}, \\
C_{d=4} \equiv \gamma^1\gamma^3, & \quad &
C^{-1}_{d=4} \left(- \gamma^{ij}\right)^{T} C_{d=4} = \gamma^{ij},  
\end{eqnarray}
we have four 4-component spinors\footnote{$c$ of $\Psi^c$ means that
$\Psi^c$ is a charge conjugation of $\Psi$.}
\begin{eqnarray}
\Psi \equiv \left(\begin{array}{c}
	\psi_\alpha \\ \bar{\chi}^{~\dot{\alpha}}
	     \end{array}\right),
& \quad &
\Psi^c \equiv B^{-1}_{d=4} \Psi^* = \left(\begin{array}{c}
			       \chi_\alpha \\ -\bar{\psi}^{\dot{\alpha}}
				    \end{array}\right), \\
\bar{\Psi} \equiv \Psi^{\dagger}(-i\gamma^0) = \left(\begin{array}{cc}
               \chi^{~\alpha}& \bar{\psi}_{\dot{\alpha}}
						     \end{array}\right),
& \quad &
\bar{\Psi}_c \equiv \Psi^T C^T_{d=4} = \barr{(\Psi^c)} =\left(\begin{array}{cc}
	     -\psi^\alpha & \bar{\chi}_{\dot{\alpha}}\\
						     \end{array}\right),
\end{eqnarray}
which transform as
\begin{eqnarray}
 \Psi \longrightarrow g \cdot \Psi, & \quad & 
 \Psi^c \longrightarrow g \cdot \Psi^c,      \nonumber \\
 \bar{\Psi} \longrightarrow \bar{\Psi} \cdot g^{-1}, & \quad &
 \bar{\Psi}_c \longrightarrow \bar{\Psi}_c \cdot g^{-1} ,
\label{eq:spinor-transf-d4}
\end{eqnarray}
where $g$ is the spinor representation of $\SO(3,1)$.
We define the chirality operator $\gamma_{(4)} \equiv (-i
\gamma_{d=4}^{0123}) = \diag(1,1,-1,-1)$.\footnote{We frequently use 
the abbreviated notation such as $\gamma^{0123} \equiv 
\gamma^0 \gamma^1 \gamma^2 \gamma^3$.}
Spinor on which $\gamma_{(4)}$ is +1 is referred to as (${\bf 2,1}$)
representation and spinor on which $\gamma_{(4)}$ is $-1$ as (${\bf 1,2}$)
representation. 
The Majorana condition is given by 
\begin{equation}
\Psi = \gamma_{(4)} \Psi^c  \quad \left( {\rm equivalent~to~} 
\Psi = \left( \begin{array}{c}
	\psi_\alpha \\ \bar{\psi}^{\dot{\alpha}}
	      \end{array}\right)\right).
\end{equation}

\subsection{The $\SO(5,1)$ gamma matrices and spinors}
6-dimensional gamma matrices $\Gamma_{d=6}^\mu~(\mu=0,1,..., 5)$ 
are given by
\begin{equation}
\Gamma^i_{d=6}=\gamma_{d=4}^i \otimes \gamma_{(2)} 
\quad(i=0,1,2,3),\quad
\Gamma^m_{d=6}={\bf 1}_{ 4 \times 4} \otimes \gamma_{(2)}^m 
\quad (m=4,5),
\label{eq:gamma-SO(5,1)}
\end{equation}
where $\gamma_{(2)}^m \equiv (\tau^2,\tau^1)$ and $\gamma_{(2)} \equiv i 
\gamma_{(2)}^{45} = \tau^3$.
Using $(4 \times 4)\otimes (2 \times 2)$ matrices $B_{d=6}$ and
$C_{d=6}$
\begin{eqnarray}
 B_{d=6} \equiv B_{d=4} \otimes \tau^1 = - \Gamma_{d=6}^{0134}, & \quad &
 B_{d=6}^{-1}\left(\Gamma^{\mu\nu}\right)^* B_{d=6} = \Gamma^{\mu\nu}, \\
 C_{d=6} \equiv C_{d=4} \otimes i \tau^2 = i \Gamma_{d=6}^{134}, & \quad &
 C_{d=6}^{-1} \left(- \Gamma^{\mu\nu}\right)^{T} C_{d=6} = \Gamma^{\mu\nu}, 
\end{eqnarray}
we have four 8-component spinors (in $4 \times 2$ matrix form)
\begin{eqnarray}
\Psi \equiv \left( \begin{array}{cc}
       \psi_\alpha & \lambda_\alpha \\ 
       \bar{\chi}^{\dot{\alpha}} & \bar{\kappa}^{\dot{\alpha}}
		   \end{array}\right),    & \quad &
\Psi^c \equiv B_{d=6}^{-1} \Psi^* = \left( \begin{array}{cc}
      \kappa & \chi \\
      - \bar{\lambda} & -\bar{\psi}
					   \end{array}\right),  \nonumber \\
\bar{\Psi} \equiv \Psi^\dagger (-i \Gamma_{d=6}^0) =\left( \begin{array}{cc}
     \chi^\alpha & \bar{\psi}_{\dot{\alpha}} \\
     -\kappa^\alpha & -\bar{\lambda}_{\dot{\alpha}}
				\end{array}\right), & \quad &
\bar{\Psi}_c \equiv \Psi^T C_{d=6}^T = \barr{(\Psi^c)}=\left(\begin{array}{cc}
    -\lambda  & \bar{\kappa} \\ 
    \psi & - \bar{\chi}
						       \end{array}\right),
\label{eq:spinor-d6}
\end{eqnarray}
which transform the same as in eq.(\ref{eq:spinor-transf-d4}) under 
the $\SO(5,1)$.
We define the chirality operator $\Gamma_{d=6}^{012345}= (\gamma_{(4)}
\otimes \gamma_{(2)})$. The chirality +1 part of a spinor is referred to as
${\bf 4}_+$ representation ({\it i.e.} $\psi$ and $\bar{\kappa}$ of $\Psi$ 
in eq.(\ref{eq:spinor-d6})), and $-1$ part as ${\bf 4}_-$ representation
({\it i.e.} $\bar{\chi}$ and $\lambda$). Since ${\bf 4}_\pm^* \simeq {\bf
4}_\pm$, the pseudo-Majorana condition can be imposed separately upon ${\bf
4}_+$ spinors $\{\Psi_{{\bf 4}_+, A}\}_{A=1,2,\cdots,2 N_+}$ or 
${\bf 4}_-$ spinors $\{ \Psi_{{\bf 4}_-, B}\}_{B=1,2,\cdots, 2 N_-}$:
\begin{equation}
 \Psi_{{\bf 4}_+,A} = {\bf J}_{AC} \Psi^c_{{\bf 4}_+,C}, \quad 
 \Psi_{{\bf 4}_-,B} = {\bf J}_{BD} \Psi^c_{{\bf 4}_-,D}, \quad 
 {\rm where ~~} {\bf J}_{AC} = {\bf J}_{BD} = \left( \begin{array}{cc}
				& {\bf 1} \\ -{\bf 1} & 
				      \end{array}\right).  
\label{eq:pseudo-majorana-d6}
\end{equation}
Or equivalently,
\begin{eqnarray}
 \Psi_{{\bf 4}_+,A} = \left(\begin{array}{cc}
		      \psi_A & \\ & -\bar{\psi}^{A+N_+}
			   \end{array}\right),    & \quad &
 \Psi_{{\bf 4}_+,A+N_+} = \left(\begin{array}{cc}
		      \psi_{A+N_+} & \\ & \bar{\psi}^{A}
			   \end{array}\right),     \\
 \Psi_{{\bf 4}_-,B} = \left(\begin{array}{cc}
		         & \chi_{B+N_-} \\ \bar{\chi}^B & 
			   \end{array}\right),    & \quad &
 \Psi_{{\bf 4}_-,B+N_-} = \left(\begin{array}{cc}
		         & -\chi_{B}\\ \bar{\chi}^{B+N_-} &
			   \end{array}\right).
\end{eqnarray}

\subsection{The $\SO(9,1)$ gamma matrices and spinors}
10-dimensional gamma matrices $\Gamma_{d=10}^\mu~(\mu=0,1,...,9)$ 
are given by
\begin{equation}
\Gamma^i_{d=10}=\gamma_{(4)}^i \otimes \gamma_{(6)} 
\quad(i=0,1,2,3),\quad
\Gamma^m_{d=10}={\bf 1}_{ 4 \times 4} \otimes \gamma_{(6)}^m 
\quad (m=4,\ldots,9),
\end{equation}
where
\begin{eqnarray}
\gamma_{(6)}^4 = \tau^2 \otimes {\bf 1} \otimes \tau^2, ~~~&
\gamma_{(6)}^6 = \tau^3 \otimes \tau^2 \otimes \tau^2, ~~~&
\gamma_{(6)}^8 = \tau^1 \otimes \tau^2 \otimes \tau^2,  \nonumber \\
\gamma_{(6)}^5 = -\tau^2 \otimes \tau^3 \otimes \tau^1, ~~~&
\gamma_{(6)}^7 = -{\bf 1} \otimes \tau^2 \otimes \tau^1, ~~~&
\gamma_{(6)}^9 = -\tau^2 \otimes \tau^1 \otimes \tau^1,
\label{eq:gamma-SO(6)}
\end{eqnarray}
and 
\begin{equation}
\gamma_{(6)} \equiv i \gamma_{(6)}^{456789} = 
{\bf 1} \otimes {\bf 1} \otimes \tau^3.
\label{eq:gamma-6}
\end{equation}
Using $(4 \times 4) \otimes (8 \times 8)$ matrices $B_{d=10}$ and
$C_{d=10}$ 
\begin{eqnarray}
 B_{d=10} \equiv B_{d=4} \otimes ({\bf 1} \otimes {\bf 1} \otimes -i \tau^2) 
          =  \Gamma_{d=10}^{013579}, & \quad &
 B_{d=10}^{-1}\left(\Gamma^{\mu\nu}\right)^* B_{d=10} = \Gamma^{\mu\nu}, \\
 C_{d=10} \equiv C_{d=4} \otimes ({\bf 1} \otimes {\bf 1} \otimes \tau^1) 
          = - i \Gamma_{d=10}^{13579}, & \quad &
 C_{d=10}^{-1} \left(- \Gamma^{\mu\nu}\right)^{T} C_{d=10} = \Gamma^{\mu\nu}, 
\end{eqnarray}
we have four 32-component spinors (in $ 4 \times 8$ matrix form) 
\begin{eqnarray}
\Psi \equiv \left( \begin{array}{cc}
       \psi_\alpha^a & \lambda_{\alpha,a} \\ 
       \bar{\chi}^{\dot{\alpha},a} & \bar{\kappa}^{\dot{\alpha}}_a
		   \end{array}\right), & \quad &
\Psi^c \equiv B_{d=10}^{-1} \Psi^* = \left( \begin{array}{cc}
      \kappa^a & -\chi_a \\
      - \bar{\lambda}^a & \bar{\psi}_a
					   \end{array}\right),  \nonumber \\
\bar{\Psi} \equiv \Psi^\dagger (-i \Gamma_{d=10}^0) =\left( \begin{array}{cc}
     \chi^\alpha_a & \bar{\psi}_{\dot{\alpha},a} \\
     -\kappa^{\alpha,a} & -\bar{\lambda}_{\dot{\alpha}}^a
				\end{array}\right), & \quad &
\bar{\Psi}_c \equiv \Psi^T C_{d=6}^T = \barr{(\Psi^c)}=\left(\begin{array}{cc}
    - \lambda_a  & \bar{\kappa}_a \\ 
    - \psi^a & \bar{\chi}^a
						       \end{array}\right),
\label{eq:spinor-d10}
\end{eqnarray}
$(a=1,2,3,4)$ which transform the same as in
eq.(\ref{eq:spinor-transf-d4}) under the $\SO(9,1)$.
We define the chirality operator $\Gamma_{d=10}\equiv
\Gamma_{d=10}^{0123456789} = (\gamma_{(4)} \otimes \gamma_{(6)})$. 
The chirality +1 part of a spinor is referred to as ${\bf 16}_+$
representation ({\it i.e.} $\psi^a$ and $\bar{\kappa}_a$ of $\Psi$ in 
eq.(\ref{eq:spinor-d10})), and $-1$ part as ${\bf 16}_-$ representation 
({\it i.e.} $\bar{\chi}^a$ and $\lambda_a$). Since ${\bf 16}_\pm^*
\simeq {\bf 16}_\pm$, the Majorana condition can be imposed separately upon 
${\bf 16}_+$ spinor $\Psi_{{\bf 16}_+}$ or ${\bf 16}_-$ spinor 
$\Psi_{{\bf 16}_-}$:
\begin{equation}
 \Psi_{{\bf 16}_\pm} = \Psi_{{\bf 16}_\pm}^c,
\label{eq:majorana-d10}
\end{equation}
or equivalently,
\begin{equation}
  \Psi_{{\bf 16}_+} = \left( \begin{array}{cc}
		      \psi_\alpha^a & \\ & \bar{\psi}^{\dot{\alpha}}_a
			    \end{array}\right), \quad
 \Psi_{{\bf 16}_-} = \left( \begin{array}{cc}
		        & -\chi_{\alpha,a} \\ \bar{\chi}^{\dot{\alpha},a} &
			   \end{array}\right).
\end{equation}

\subsection{Reduction from $d=10$ to $d=6$}
Let us consider the action of gamma matrices
$\Gamma_{d=10}^{0,1,\cdots,5}$ on 32-component spinors.

Since the 2nd factor of $\gamma_{(6)}^{4,5}$ (in
eq.(\ref{eq:gamma-SO(6)})) and $\gamma_{(6)}$ (in eq.(\ref{eq:gamma-6})) 
are all ${\bf 1}$ or $\tau^3$, the $\SO(5,1)$ action does not mix 
\begin{equation}
 \left( \begin{array}{cc}
  \psi^a & \lambda_a \\ \bar{\chi}^a & \bar{\kappa}_a
	\end{array}\right)_{a=1,2} \quad {\rm and } \quad 
  \left( \begin{array}{cc}
   \psi^a & \lambda_a \\ \bar{\chi}^a & \bar{\kappa}_a
	 \end{array}\right)_{a=3,4}.
\end{equation}
On each part ($a=1,2$ and $a=3,4$),
\begin{eqnarray}
\Gamma_{d=10}^4 = {\bf 1}_{4 \times 4} \otimes \left(\tau^2 \otimes \gamma_{(2)}^4\right), & \quad & 
\Gamma_{d=10}^4 = {\bf 1}_{4 \times 4} \otimes \left(\tau^2 \otimes \gamma_{(2)}^4\right), \\
\Gamma_{d=10}^5 = {\bf 1}_{4 \times 4} \otimes \left(-\tau^2 \otimes \gamma_{(2)}^5\right), & \quad & 
\Gamma_{d=10}^5 = {\bf 1}_{4 \times 4} \otimes \left(\tau^2 \otimes \gamma_{(2)}^5\right). 
\end{eqnarray}
If we rearrange the basis of spinors as follows:
\begin{equation}
\Psi_{{\bf 8},A} = \left(\begin{array}{cc}
		   i \lambda_a & \epsilon_{ae}\psi^e \\
                   - \bar{\kappa}_a & i \epsilon_{ae} \bar{\chi}^e
			\end{array}\right)_{e,a=A=1,2}, \quad
\Psi_{{\bf 8},B} =\left(\begin{array}{cc}
		  -\epsilon_{bd} \psi^d & i \lambda_b \\
                  i \epsilon_{bd} \bar{\chi}^d & \bar{\kappa}_b
		       \end{array}\right)_{d,b=B+2=3,4},
\end{equation}
then the gamma matrices are now the same as those given in 
eq.(\ref{eq:gamma-SO(5,1)}) on each $\Psi_{{\bf 8},A}$($A=1,2$) or 
$\Psi_{{\bf 8},B}$($B=1,2$). We can see that the 10-dimensional 
Majorana condition in eq.(\ref{eq:majorana-d10}) is equivalent to the 
6-dimensional pseudo-Majorana conditions on $\{\Psi_{{\bf 8},A}\}_{A=1,2}$ 
and $\{\Psi_{{\bf 8},B}\}_{B=1,2}$ in eq.(\ref{eq:pseudo-majorana-d6}).

\section{SUSY breaking due to the presence of a 3-brane}
Algebra of 6-dimensional (0,1) SUSY is given by 
\begin{eqnarray}
\{ {\cal Q}^{(6)}_{{\bf 4}_-,B}, {\cal Q}^{(6)}_{{\bf 4}_-,D}\}
&=& -i \epsilon_{BD} \left(\Gamma_{d=6}^\mu C_{d=6}^{T,-1}\right)P_\mu  
\nonumber \\
   && + \tau^{a ~~C}_B \epsilon_{CD} 
   \left(\Gamma^{\lambda\mu\nu}_{d=6}C_{d=6}^{T,-1}\right)C^a_{\lambda\mu\nu}
  + i \epsilon_{BD} \left(\Gamma_{d=6}^{\mu\nu\rho\sigma\tau} 
      C_{d=6}^{T,-1}\right)C_{\mu\nu\rho\sigma\tau} ,
\end{eqnarray}
 where the $C^a_{\lambda\mu\nu}(a=1,2,3)$ in the 2nd term and
$C_{\mu\nu\rho\sigma\tau}$ in the 3rd term in the right hand side 
 are charges of objects that extend in 3 and 5 spatial dimensions, that
is, charges of 3-brane and 5-brane (6-dimensional bulk itself), 
respectively. Let us consider the case where a 3-brane exists and see
 how the unbroken SUSY charge is determined. 
Suppose that $\vev{C^3_{123}} \neq 0$. (This corresponds to the 
assumption that the
3-brane stretches in 123 space directions.\footnote{
$\tau_{B}^{a ~~C}\vev{C^a_{123}} \neq 0$ breaks $\SU(2)_{{\bf 4}_-}$
 R-symmetry. When the D5-branes (in the 10-dimensional picture) wrap into 89
 directions, the breaking of the $\SO(4)_{6789} \simeq \SU(2)_{{\bf
 4}_-} \times \SU(2)_{{\bf 4}_+}$ is precisely given by 
 $\vev{C^3_{123}} \neq 0$. }) 
Reduction of this algebra into that of 4-dimensional effective theory is
given by 
\begin{eqnarray}
\{ {\cal Q}^{(4)2}_\alpha ,\bar{\cal Q}^{(4)}_{\dot{\beta},2} \} &=& 
          i (i \sigma^i)_{\alpha\dot{\beta}}P_i 
          - (i\sigma^1 i\bar{\sigma}^2 i \sigma^3)_{\alpha\dot{\beta}} 
                  C^3_{123}, \label{eq:6-4algebra2}\\
\{ {\cal Q}^{(4)1}_\alpha ,\bar{\cal Q}^{(4)}_{\dot{\beta},1}\} &=& 
           i (i \sigma^i)_{\alpha\dot{\beta}}P_i 
          + (i\sigma^1 i\bar{\sigma}^2 i\sigma^3)_{\alpha\dot{\beta}}
                            C^3_{123}, \label{eq:6-4algebra1}\\
 \{ {\cal Q}^{(4)1}_\alpha , {\cal Q}^{(4)2}_\beta \}&=&
                            \epsilon_{\alpha\beta} (-P_4 + i P_5).
\end{eqnarray}
Then it is easily seen that  
${\cal Q}^{(4)1}$ is left unbroken (eq.(\ref{eq:6-4algebra1})) 
while ${\cal Q}^{(4)2}$ not (eq.(\ref{eq:6-4algebra2})),
if the 3-brane tension $\vev{P^0}$ is related to the 3-brane charge 
$\vev{C^3_{123}}$ as $\vev{P^0} = - \vev{P_0} =
\vev{C^3_{123}}$ ({\it i.e. }BPS condition).\footnote{
Although the momenta $P_4$ and $P_5$ are not well-defined since the
3-brane breaks these symmetries, they do not appear in the SUSY algebra 
of the unbroken ${\cal Q}^{(4)1}$.} 
Otherwise, both ${\cal Q}^{(4)1}$ and  ${\cal Q}^{(4)2}$ are broken.

The same argument as above holds also in the 6-dimensional (1,0) SUSY.
Therefore, to keep the ${\cal N}=2$ SUSY on the 3-brane, we need (1,1)
SUSY or (2,0) SUSY in the 6-dimensional bulk, and each 6-dimensional
SUSY charge provides one 4-dimensional SUSY charge, if the 3-brane
satisfies the BPS condition. 

\section{Extended SUSY transformation in 4-dimensional gauge 
theories}
When a 4-dimensional effective theory is derived from a 10-dimensional
theory through a compactification, the $\SO(6)_{456789}$ subgroup is
regarded as an internal symmetry. 
The $(4 \times 8)$-matrix expression of the 32-component spinor is now
regarded as eight 4-component $\SO(3,1)$ spinors, and these eight
spinors form a ${\bf 4}+{\bf 4}^*$ multiplet under the internal symmetry 
$\SU(4)_{\rm R}$.
${\bf 4}^*$ representation of the $\SU(4)_{\rm R}$ is
simply given by complex conjugate of ${\bf 4}$ with the convention 
 of the $\SO(6)_{456789}$ gamma matrices in eq.(\ref{eq:gamma-SO(6)}).
Explicit representation of the $\SU(4)_{\rm R}$ can be 
calculated from eq.(\ref{eq:gamma-SO(6)}). For example, generators of
the Cartan subalgebra are given by 
\begin{equation}
\frac{i}{2}\gamma_{(6)}^{45}=
            \frac{-1}{2} {\bf 1} \otimes \tau^3 \otimes \tau^3, \quad
\frac{i}{2}\gamma_{(6)}^{67}=
            \frac{-1}{2} \tau^3 \otimes {\bf 1} \otimes \tau^3, \quad
\frac{i}{2}\gamma_{(6)}^{89}=
            \frac{-1}{2} \tau^3 \otimes \tau^3 \otimes \tau^3. 
\end{equation} 

SUSY charge in 10-dimensional SUSY Yang-Mills theories and its
infinitesimal transformation parameter are 
\begin{equation}
 {\cal Q}_{{\bf 16}_+}^{(10)} = \left(\begin{array}{cc}
	 {\cal Q}^{(4),a}_\alpha & \\ & \bar{{\cal Q}}^{(4)\dot{\alpha}}_a
				     \end{array}\right) \quad
 \epsilon_{{\bf 16}_-} = \left( \begin{array}{cc}
	     & -\epsilon_{\alpha,a}\\ \bar{\epsilon}^{\dot{\alpha},a} & 
		   \end{array}\right).
\end{equation} 
Infinitesimal SUSY transformation is decomposed into 4-dimensional
spinors as 
\begin{equation}
 \bar{\epsilon}_c {\cal Q}= \left( \begin{array}{cc}
			     \epsilon_a & \\ & \bar{\epsilon}^a
				   \end{array}\right)
\left( \begin{array}{cc}
  {\cal Q}^{(4),a} & \\ & \bar{{\cal Q}}^{(4)}_a
       \end{array}\right)
\rightarrow \epsilon_a {\cal Q}^{(4),a} + 
            \bar{\epsilon}^a \bar{{\cal Q}}^{(4)}_a.
\end{equation}
Here, $\bar{\epsilon}_c$ in the left-hand side is a 10-dimensional
spinor defined as in eq.(\ref{eq:spinor-d10}).
The 4-dimensional SUSY charges ${\cal Q}^{(4),a}$ transform under 
the $\SU(4)_{\rm R}$ as ${\bf 4}$ and $\bar{{\cal Q}}^{(4)}_a$ as 
${\bf 4}^*$. Transformation parameters $\epsilon_a$ transform as 
${\bf 4}^*$ and $\bar{\epsilon}^a$ as ${\bf 4}$.
Gauge fermions in the 10-dimensional Yang-Mills theory are described by
${\bf 16}_-$ spinor, 
\begin{equation}
 \Lambda_{{\bf 16}_-} = \left(\begin{array}{cc}
			 & - \chi_{\alpha,a}\\ \bar{\chi}^{\dot{\alpha},a} &
			      \end{array}\right), 
\end{equation}
where $\chi_a$ transform under the $\SU(4)_{\rm R}$ as ${\bf 4}^*$.

SUSY transformation of the 10-dimensional SUSY Yang-Mills theory is given by
\begin{eqnarray}
\delta A_\mu = \bar{\epsilon}_c \Gamma_{d=10,\mu} \Lambda, \\
\delta \Lambda = \frac{1}{2}\Gamma_{d=10}^{\mu\nu}F_{\mu\nu}\epsilon ,
\label{eq:d10-SUSYtransf-fermion}
\end{eqnarray}
whose decomposition into 4-dimension ({\it i.e.} 4-dimensional ${\cal
N}=4$ SUSY transformation) is
\begin{eqnarray}
\delta A_i &=& i \left( \epsilon_a \sigma_i \bar{\chi}^a 
                    - \chi_a \sigma_i \bar{\epsilon}^a \right),  \\
\delta \sigma_{ab} &=& \sqrt{2} \{ (\epsilon_a \chi_b - \epsilon_b \chi_a )
                           - \epsilon_{abcd} \bar{\epsilon}^c \bar{\chi}^d\},\\
\delta \chi_a &=& \frac{1}{2}\gamma_{d=4}^{ij}\epsilon_a F_{ij}
                  +  \sqrt{2} i \sigma^{i}\bar{\epsilon}^b (D_i \sigma_{ab})
                  - i \left[ \sigma_{ab},\sigma^{\dagger,bc}\right]\epsilon_c.
\label{eq:N4-SUSYtransf-fermion}
\end{eqnarray}
Here, $\epsilon_{abcd}$ is the totally anti-symmetric tensor, $D_i$ the
covariant derivatives, and $\sigma_{ab}$ is defined as 
eq.(\ref{eq:N4-scalar-multiplet}), where $\sigma \equiv 
(1/\sqrt{2})(A_4 + i A_5)$, $\sigma' \equiv (1/\sqrt{2})(A_6 + i A_7)$ 
and $\sigma'' \equiv (1/\sqrt{2})(A_8 + i A_9)$. 
The 1st term in the right-hand side of
eq.(\ref{eq:N4-SUSYtransf-fermion}) comes from 
$\Gamma^{ij}F_{ij}$ of eq.(\ref{eq:d10-SUSYtransf-fermion}), the 2nd term 
of eq.(\ref{eq:N4-SUSYtransf-fermion}) from $\Gamma^{in}F_{in}$ of 
 eq.(\ref{eq:d10-SUSYtransf-fermion}) and the 3rd term from
$\Gamma^{mn}F_{mn}$ ($i,j = 0,1,2,3$, $m,n = 4,..., 9$).
If there exists a background flux $\vev{F_{mn}} \ne 0$, then another
contribution arises from the $\Gamma^{mn}F_{mn}$ in 
eq.(\ref{eq:N4-SUSYtransf-fermion}). This contribution is given by
\begin{equation}
 \delta \Lambda =  \frac{1}{2}\Gamma^{mn}\vev{F_{mn}} \epsilon;
\end{equation}
that is,
\begin{equation}
 \delta \left( \begin{array}{c}
  \chi_1 \\ \chi_2 \\ \chi_3 \\ \chi_4 \\
	\end{array}\right) = \frac{i}{2} \left( \begin{array}{cccc}
 -\eta''-\eta-\eta' & 2 \eta_+ & 2 \eta'_+ & 2 \eta''_+ \\
 2 \eta_- & -\bar{\eta}''+\eta+\bar{\eta}' & 2 \bar{\eta}''_+ & 
2 \bar{\eta}'_-  \\
 2 \eta'_- & 2 \bar{\eta}''_- & \bar{\eta}''-\bar{\eta}+\eta' &
 2  \bar{\eta}_+  \\
 2 \eta''_- & 2 \bar{\eta}'_+ & 2 \bar{\eta}_- & \eta'' + 
      \bar{\eta} - \bar{\eta}' 
					\end{array}\right) 
   \left(\begin{array}{c}
    \epsilon_1 \\ \epsilon_2 \\ \epsilon_3 \\ \epsilon_4 \\
	 \end{array}\right),
\end{equation}
where
\begin{eqnarray}
\eta \equiv \eta^3 \quad \eta_\pm \equiv \eta^1 \pm i \eta^2 \quad 
 \eta^a \equiv \eta^{a,mn} \vev{F_{mn}} \quad (a = 1,2,3; m,n =6789), \\
\eta' \equiv \eta^{'3} \quad \eta'_\pm \equiv \eta^{'1} \pm i \eta^{'2} \quad 
 \eta^{'a} \equiv \eta^{a,mn} \vev{F_{mn}} \quad (a = 1,2,3; m,n =8945), \\
\eta'' \equiv \eta^{''3} \quad \eta''_\pm \equiv \eta^{''1} \pm i \eta^{''2} 
\quad 
 \eta^{''a} \equiv \eta^{a,mn} \vev{F_{mn}} \quad (a = 1,2,3; m,n =4567).
\end{eqnarray}
Here, $\eta^{a,mn}$ is the 'tHooft symbol \cite{tHooft}.
$\bar{\eta},\bar{\eta}',\bar{\eta}'',\bar{\eta}_{\pm},\bar{\eta}'_{\pm},
\bar{\eta}''_{\pm}$ are also defined in an analogous way, using the 
'tHooft symbol $\bar{\eta}^{a,mn}$.


The 4-dimensional ${\cal N}=2$ SUSY transformation (transformation parameter 
$(\delta\theta)_A$) of fields in an ${\cal N}=2$ vector multiplet 
(scalar $\phi \equiv i\sigma$, gaugino $\psi^A$ and vector $A_i$) 
is given by
\begin{eqnarray}
\delta A_i &=& i\left( (\delta \theta)_A \sigma_i \bar{\psi}_B \epsilon^{AB} - 
                    \psi^A \sigma_i \barr{(\delta \theta)}^B \epsilon_{AB}\right) \\
\delta \phi &=& \sqrt{2} (\delta \theta)_A \psi^A \\
\delta \psi^A &=& \frac{-1}{2}\gamma^{ij}_{d=4}F_{ij}\epsilon^{AB}(\delta \theta)_B 
                - \sqrt{2} i \sigma^i \barr{(\delta \theta)}^A (D_i \phi)
                + i \left[ \phi , \phi^\dagger \right] \epsilon^{AB} (\delta \theta)_B 
       \nonumber \\
                && -i \epsilon^{AB} \left( \begin{array}{cc}
                      D & -\sqrt{2}i F^* \\ \sqrt{2}i F & -D  
                           \end{array}\right)_{B}^{~~~~C}(\delta \theta)_C
      \nonumber \\
             &&  +i \epsilon^{AB} \left( \begin{array}{cc}
		    \frac{\eta''+\eta+\eta'}{2} &  i \eta''_+ \\
                    -i \eta''_- & - \frac{\eta''+\bar{\eta}-\bar{\eta}'}{2}
                          \end{array}\right)_B^{~~~~C} (\delta \theta)_C , 
\end{eqnarray}
where $A,B,C$ denote indices of the $\SU(2)_{\rm R}$ doublets, $\psi^A
= (\chi_4,i\chi_1)$, $(\delta \theta)_A = (i\epsilon_1, -\epsilon_4)$
and $(-\sqrt{2}\Im F,\sqrt{2}\Re F,D)$ the $\SU(2)_{\rm R}$ triplet 
auxiliary fields. 
If $\vev{F_{mn}}\ne 0$ for $m,n=4567$ while other $\vev{F_{mn}}$ are
zero, then $\eta+\eta'=\eta''$ and $\bar{\eta}-\bar{\eta}'=\eta''$ follow.
We can see that the $\eta''$($\eta''_\pm$), or equivalently 
the $\vev{F_{45}+F_{67}}$ ($\vev{F_{56}+F_{47} \pm i (F_{64}+F_{57})}$), 
play the role of the Fayet-Iliopoulos D-(F-) term parameters in the 
${\cal N}=2$ SUSY gauge theories. 

\section{Unbroken SUSY condition from orientifold planes}

An orientifold is an orbifold where the ${\bf Z}_2$-flip orbifold group 
action is accompanied by a flipping between left-mover and right-mover 
on the string worldsheets.
This left-right flipping changes the SUSY generator ${\cal Q}^{(10)}_{{\bf
16}_+}$ into ${\cal Q}^{(10)c}_{{\bf 16}_+}$ and vice versa, and this is 
a crucial difference from the ordinary orbifold.
Therefore, the unbroken SUSY condition from the O9-plane is given by
\begin{equation}
 {\cal Q}^{(10)} = {\cal Q}^{(10)c},
\end{equation}   
and the condition from the O5-planes by
\begin{equation}
 {\cal Q}^{(10)} = 
\exp \left(i \pi (T^{45}_{(6)}-T^{67}_{(6)})\right) {\cal Q}^{(10)c}
 = -\Gamma_{d=10}^{4567} {\cal Q}^{(10)c}.
\end{equation}   
Note that ${\cal Q}^{(10)c}$, not ${\cal Q}^{(10)}$, appears in the
right-hand sides, contrary to the the case of ordinary orbifolding. 
Remembering that the SUSY charge satisfies the chirality
condition ${\cal Q}^{(10)}_{{\bf 16}_+} = \Gamma_{d=10} {\cal Q}^{(10)}_{{\bf
16}_+}$ and ${\cal Q}^{(10)c}_{{\bf 16}_+} = \Gamma_{d=10} {\cal
Q}^{(10)c}_{{\bf 16}_+}$, we can easily see that the above two
conditions are equivalent to the unbroken SUSY conditions arising from 
D9-branes (eq.(\ref{realitycond})) and D5-branes (eq.(\ref{d5susy})).
Therefore, the ``${\bf Z}_2$'' orbifoldings associated with the
orientifold planes parallel to the D-branes do not cause further
breakings of the SUSY. 

%
%
%
%
\newcommand{\Journal}[4]{{#1} {\bf #2}, {#3} {(#4)}} 

\begin{table}[h]
\begin{center}
\begin{tabular}{c||ccc|cc||c|cc|cc}
 & ${\bf 5}^*$ & ${\bf 10}$ & ${\bf 1}$   & $H$   & $\bar{H}$  
 & $X^\alpha_\beta,X_0$  & $Q^k$      & $\bar{Q}_k$ & $Q^6$ & $\bar{Q}_6$ \\
\hline 
${\bf Z}_{4 {\rm R}}$ charge 
 & 1          & 1           & 1           & 0     & 0
 & 2          & 0           & 0           & 2     & $-2$ \\ 
\end{tabular}
\caption{\label{tab:R-charge} Charge assignment of the discrete
 ${\bf Z}_{4{\rm R}}$ symmetry}
\end{center}
\end{table} 

\begin{table}[htb]
\begin{center}
\begin{tabular}{cccc}
& $\SO(2)_{45}$ & $\SO(2)_{67}$ & $\SO(2)_{89}$ \\
\hline
${\cal Q}^{(4)1}$ & $-1/2$ & $-1/2$ & $-1/2$ \\
\hline
$\chi_1^{\U(5)_{\rm GUT}}$, $\chi_1^{\U(3)_{\rm H}}$ & 1/2 & 1/2 & 1/2 \\
$A_\mu^{\U(5)_{\rm GUT}}$, $A_\mu^{\U(3)_{\rm H}}$ & 0 & 0 & 0 \\
$\sigma$, $x$ & 1 & 0 & 0 \\
$\sigma'$, $x'$ & 0 & 1 & 0 \\
$\sigma''$, $x''$ & 0 & 0 & 1 \\
$\chi_2$, $\tilde{x}$ & 1/2 & $-1/2$ & $-1/2$ \\
$\chi_3$, $\tilde{x}'$ & $-1/2$ & 1/2 & $-1/2$ \\
$\chi_4$, $\tilde{x}''$ & $-1/2$ & $-1/2$ & 1/2 \\
$q^k$, $\overline q_k$ & 1/2 & 1/2 & 0 \\
$\tilde{q}^k$, $\tilde{\bar{q}}_k$ & 0 & 0 & $-1/2$ \\
\end{tabular}
\caption{The field contents arising from open strings
 on the D3-D7 brane system transform under the space
 rotational symmetry $\SO(6)_{456789}$. Charges of these fields for the
maximal torus of this  $\SO(6)_{456789}$ ({\it i.e.}
 $\SO(2)_{45} \times \SO(2)_{67} \times \SO(2)_{89}$ ) are summarized. 
Charges are determined by observing which open-string modes
correspond to those fields.
They are combined into chiral multiplets of ${\cal N}=1$ SUSY
 of ${\cal Q}^{(4)1}$ as
${\cal W}_{\alpha}^{\U(5)_{\rm GUT}} = (\chi_1^{\U(5)_{\rm GUT}},
F_{ij}^{\U(5)_{\rm GUT}})$,
${\cal W}_{\alpha}^{\U(3)_{\rm H}} = (\chi_1^{\U(3)_{\rm H}},
F_{ij}^{\U(3)_{\rm H}})$,
$\Sigma=(\sigma,\chi_2)$, $X=(x,\tilde{x})$,
$\Sigma'=(\sigma',\chi_3)$, $X'=(x',\tilde{x}')$,
$\Sigma''=(\sigma'',\chi_4)$, $X''=(x'',\tilde{x}'')$,
$Q^k=(q^k,\tilde{q}^k)$ and $\bar{Q}_k=(\bar{q}_k,\tilde{\bar{q}}_k)$. Note
that the charges of the 2nd-lowest component of all chiral multiplets
are $-1/2$ smaller than those of the lowest components. This means that 
these symmetries can be regarded as R-symmetry.}
\label{u1charges.tbl}
\end{center}
\end{table}

\end{document}